UNIVERSITÀ DI PISA

Scuola di Dottorato in Ingegneria "Leonardo da Vinci"

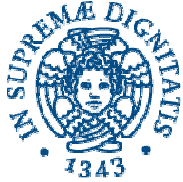

Corso di Dottorato di Ricerca in
INGEGNERIA DELL'INFORMAZIONE

Tesi di Dottorato di Ricerca

# High availability using virtualization

*Autore:*

Federico Calzolari   ______________

*Relatori:*

Prof.  Gigliola Vaglini   ______________

Prof.  Andrea Domenici ______________

Anno 2006

To Andrea, my friend

## ACKNOWLEDGMENTS

Thanks to all colleagues from Scuola Normale Superiore, National Institute for Nuclear Research INFN Pisa, European Center for Nuclear Research CERN in Geneva, for their help, support and collaboration.
Many Thanks to my tutors, Prof. Gigliola Vaglini and Prof. Andrea Domenici.
A special thanks goes to Veronica, who passed on the passion of studying to me.

# SOMMARIO

*Garantire un servizio in alta affidabilita' e' da sempre uno dei principali problemi di un centro di calcolo. Fino ad ora servizi di alta affidabilita' erano ottenuti duplicando uno ad uno i vari computer di un centro, metodo altamente dispendioso in termini di hardware e tempo uomo. Un nuovo approccio al problema puo' essere offerto dalla virtualizzazione.*

*Utilizzando la virtualizzazione e' possibile ottenere un sistema di ridondanza per tutti i servizi ospitati da un centro di calcolo. questo nuovo approccio all'alta affidabilita' permette di distribuire le macchine virtuali attive in determinato momento sui soli server attivi e disponibili, sfruttando le caratteristiche del layer di virtualizzazione: accensione, spegnimento e spostamento di macchine virtuali tra gli host fisici presenti.*

*Il sistema (3RC), basato su una macchina a stati finiti con isteresi, offre la possibilita' di far ripartire ogni macchina virtuale su un determinato host fisico, o reinstallare la stessa da zero. Una complessa infrastruttura e' stata realizzata per installare sistema operativo e middleware in pochi minuti. Al fine di virtualizzare i principali servizi di un centro di calcolo e' stata sviluppata una nuova procedura per migrare un host da fisico a virtuale.*

*L'intero nodo Grid SNS-PISA sta al momento girando in ambiente virtuale in alta affidabilita'.*

*Come estensione dell'architettura 3RC, sono state testate diverse soluzioni di storage per salvare e centralizzare tutti i dischi virtuali, da sistemi NAS a SAN, al fine di garantire la sicurezza dei dati e l'accesso da ovunque.*

*Sfruttando la virtualizzazione e la capacita' di reinstallare un computer in modo completamente automatico, viene fornita una sorta di host a richiesta, dove una qualunque azione su una macchina virtuale e' eseguita solo nel momento in cui effettivamente succede il disastro.*

# ABSTRACT

High availability has always been one of the main problems for a data center. Till now high availability was achieved by host per host redundancy, a highly expensive method in terms of hardware and human costs. A new approach to the problem can be offered by virtualization.
Using virtualization, it is possible to achieve a redundancy system for all the services running on a data center. This new approach to high availability allows to share the running virtual machines over the servers up and running, by exploiting the features of the virtualization layer: start, stop and move virtual machines between physical hosts.

The system (3RC) is based on a finite state machine with hysteresis, providing the possibility to restart each virtual machine over any physical host, or reinstall it from scratch. A complete infrastructure has been developed to install operating system and middleware in a few minutes. To virtualize the main servers of a data center, a new procedure has been developed to migrate physical to virtual hosts.

The whole Grid data center SNS-PISA is running at the moment in virtual environment under the high availability system. As extension of the 3RC architecture, several storage solutions have been tested to store and centralize all the virtual disks, from NAS to SAN, to grant data safety and access from everywhere.
Exploiting virtualization and ability to automatically reinstall a host, we provide a sort of host on-demand, where the action on a virtual machine is performed only when a disaster occurs.

**INDEX**

# 1 INTRODUCTION

One of the most critical issues for a computing center is the ability to provide a high availability service for all the main applications running on it. Today all the services are intended to be 24x7 (24 hours at day, 7 days per week), even those with a short life span.
Someone said "When the world can access your applications, application failures are exposed to a much wider community". High availability takes care of the strategies to reduce to the minimum the applications' downtime.
The goal of all availability systems is to maximize the uptime of the various online systems for which they are responsible, to make them completely fault tolerant. There are several approaches able to maximize availability: redundancy, reliability, repairability, recoverability, responsiveness, robustness.

The strong demand for high-availability solutions has generated a lot of design strategies to provide a reliable service. These strategies involve systems such as heartbeat and cluster computing - very reliable but at the same time very expensive solutions in terms of hardware and human cost.
The basic concept of a heartbeat system is the host per host redundancy of all the main servers of a computing center. The cluster computing solution is instead based on a middleware cluster suite distributed over a large number of nodes in a cluster; it is able to detect errors or failures on one node and automatically transfer workload or applications to another active node in the same cluster. Such systems typically have redundant hardware and software that make the applications available despite failures.

A lot of other commercial solutions exist at the moment to satisfy the needs of service availability in a production environment. The problem is often related to the low portability of the solution, due to the hard link between the high availability tool provided and the operating system, or to the need of a specific software.
On top of that, a lot of proprietary solutions are not free, with an often too high cost for a research computing center or for a small company.

While high availability services are essential for 24x7 mission critical applications, the cost issue has to be carefully evaluated. To extend this service level to above a 99.9 percent (three nines) availability, the cost increases exponentially. Because of the high costs and hardware configuration requirements, a five nines availability level often implies a strong negative return on investment.

My idea is to satisfy the research and the enterprise needs of high availability with a zero cost new solution. This new approach is based on the concept that a relaxed system may ensure the application redundancy required in the greater part of cases. By "relaxed" I mean a system able to restore any previously running application in less than ten minutes from the crash time.
By analyzing the causes of planned and unplanned downtime of a large computing center (INFN Pisa: 2000 CPU, 500 TB disk) for a period of three years, a new approach to high availability has been designed, based on hosts running on virtual environment.

A high availability virtual infrastructure ensures that a service has constant availability of network, processors, disks, memory. This way a failure of one of these components is transparent to the application, with a maximum time delay of five to ten minutes.

3RC is the name of the project, acronym for 3 Re Cycle. The originality of this new approach to high availability is that a computing center system manager does not have to worry about the system redundancy till the disaster occurs. At that moment the system is able to restore the crashed application in a location resulting of a choice algorithm.
The solution has been developed by using a finite state machine with hysteresis, where each state plays the role of an action to be performed on the crashed virtual machine or on the virtual layer running over the physical hosts.
With no additional costs in terms of hardware and software for a computing center, and light operating costs, this new approach offers the ability to guarantee a relaxed high availability level. The whole Grid data center SNS-PISA is running at the moment in a virtual environment under this new high availability system.

A complex infrastructure, based solely on servers available in a typical computing center, has been developed in order to install the operating system and the middleware in a few minutes, by exploiting Preboot Execution Environment (PXE) technology, as well as DNS and DHCP services. Using this infrastructure, more than 5000 servers have been installed from scratch starting by the only knowledge of their MAC address.
As an extension of the high availability architecture, several storage solutions, from Network Attached Storage (NAS) to Storage Area Network (SAN) have been tested to store and centralize all the virtual disks, to grant data safety and access from everywhere, enhancing the aggregate bandwidth and reducing at the same time the downtime period. A lot of possible solutions are available, depending on the required availability level, on the financial resources, on the storage redundancy method.

As a spin-off of this work, exploiting virtualization and ability to automatically reinstall a host, we could provide a sort of Host on-demand, where the user can ask for a customized (self configured) host in terms of processor number, RAM, disk space, for a given time, with administrator privileges. At the end of the scheduled time the machines are destroyed.
This way it is possible to optimize the resource sharing across a large number of users, by simply redistributing load upon all the available machines.

This thesis is organized in several chapters, the first Chapter being this "Introduction".
Chapter 2, "High availability", gives the state of the art of the high availability solutions currently present on the market, such as Heartbeat or Clustering computing.
Chapter 3, "Virtualization", presents several available virtualization solutions.
In chapter 4, "Scenario ad use case", I introduce the scenario of my work: a Grid data center, with all its needs, often very similar to other computing center needs.

Chapter, 5 "Infrastructure", describes the whole infrastructure to implement an efficient PXE environment, to automatically install and manage a large set of hosts in parallel.

Chapter 6, "Storage solutions", analyzes the available storage architectures, from cheap Network Attached Storage (NAS) to more expensive Storage Area Network (SAN).

Chapter 7, "Proposal and solutions", analyzes the methods used to implement a new and low cost high availability system using virtualization, and shows the experimental data resulting from the simulations.

Chapter 8, "Operation", introduces the measures taken in order to achieve an acceptable high availability level, showing a concrete use case.

Chapter 9, "Spin-off", shows some possible future outcomes resulting from my work, such as Host on-demand.

Chapter 10, "Conclusions", gives a synthetic point of view of this new approach to high availability, showing implications and limitations of this system.

In Appendix there is the kernel source of 3RC High availability project.

# 2 HIGH AVAILABILITY

## 2.1 Availability definition

The term availability (or reliability) refers to the probability that a system is operating properly at a given time [from eventhelix.com]. The ability of a system to operate despite the failure of one or more of its components is called fault tolerance. In my dissertation terms as fault tolerance, high availability, reliability, may be used interchangeably.

A failure is when a service does not work properly, e.g. a state of abnormal operation or, more precisely, not in accordance with specifications.

The main goal of a high availability system is to provide a service in accordance with specification despite errors. A solution to this issue is usually achieved by introducing a redundancy level. High availability includes all the measures that aim to guarantee service availability.

If it is not possible to avoid incidents, it is very useful to know the time necessary to reestablish service.

## 2.2 Availability parameters

### 2.2.1 MTBF

Mean Time Between Failures [from eventhelix.com]: the average time between failures of hardware modules. It is the average time a manufacturer estimates before a failure occurs in a hardware module. MTBF for hardware modules can be obtained from the vendor. MTBF for software can be determined by multiplying the defect rate with KLOCs (thousand of lines of code) executed per second.

Failures in time (FITS) is a more intuitive way of representing MTBF. FITS is nothing but the total number of failures of the module in a billion hours (i.e. 1.000.000.000 hours).

### 2.2.2 MTTR

Mean Time To Repair [from eventhelix.com]: the time taken to repair a failed hardware module. In an operational system, repair usually means replacing the hardware module. Thus the hardware MTTR can be viewed as mean time to replace a failed hardware module. Note that a low MTTR requirement entails a high operational cost for the system.

The estimated Hardware MTTR are: 30 minutes to 24 hours for an on site operation, 1 day to 1 week for an off site operation.

The estimated Software MTTR can be computed as the time taken to reboot after a software fault is detected: from 30 seconds to 30 minutes, if a manual operator reboot is required.

## 2.3 Availability measure

Availability is often expressed by the availability rate (percentage, or number of nines), measured by dividing the time the service is available by the total time, or to be more precise Availability = MTBF / (MTBF + MTTR).

Availability is typically specified in nines notation. An availability of 99.9% (3 nines) refers to a downtime of 8.76 hours/year.
Telecommunications devices need to be guaranteed for an efficiency of 5 nines, that is 5 minutes/year.

## 2.4 Risk evaluation

The failure of a computer system can cause losses in productivity, money and in critical cases material and human losses. It is therefore necessary to evaluate the risks linked to the failure of one or more of the components of a computer system.
There are several ways in which a computer system can fail. The principal failure causes are: physical (i.e. natural disaster, environment weather, humidity, temperature, material failure, network failure, power cut), human (intentional or accidental: design error, software bug, malicious attack), operational (linked to system state at a given moment: software failure).

## 2.5 Fault tolerance

When one of the resources breaks down, other resources can take over in order to give operational continuity to the system, and give the administrators the time to solve the problem. This technique is called Failover. In an ideal model [1], in case of material failures, the faulty devices should be hot swappable, i.e. capable of being extracted and replaced without service interruption.

Fig. 2.1 shows a typical Fault tolerant architecture.

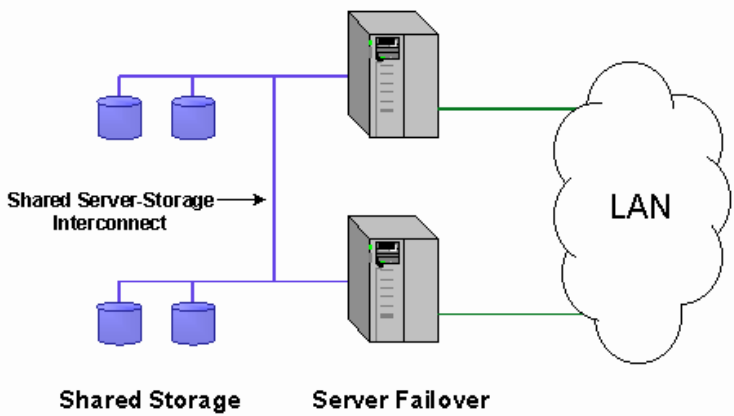

Fig. 2.1: Fault tolerant system

## 2.6 Backup

Setting up a redundant architecture and infrastructure ensures that the service will be available, but does not protect the data against errors of whatever nature: natural disasters, software corruption, bugs, or malicious attacks. Therefore it is necessary to set up backup mechanisms in order to guarantee data safety. Ideally the backup should be stored in a remote location, to avoid the possibility of corruption of both the copies in case of natural disaster.
User errors may cause a system failure or malfunctioning. In certain circumstances, a module or a system failure may not be repairable. A backup and a recovery facility infrastructure, intended to backup the system at scheduled intervals and restore a backup when a disaster occurs, can remedy this kind of problem.
A backup infrastructure can also be used for historical archives storage, to save data in a precise state, corresponding to a given time. This solution can be used for future recovery of a previous properly operational state of the system. The technical name of a backup of a whole computer system is "snapshot".
By saving a snapshot series at a scheduled time, it is possible to achieve a sort of time machine, capable of recovering the state of a computer system at a given date.

Fig. 2.2 [VMware] shows a complex storage and backup system, able to provide a centralized storage for the virtual machines, and a full backup of the virtual machines snapshots.

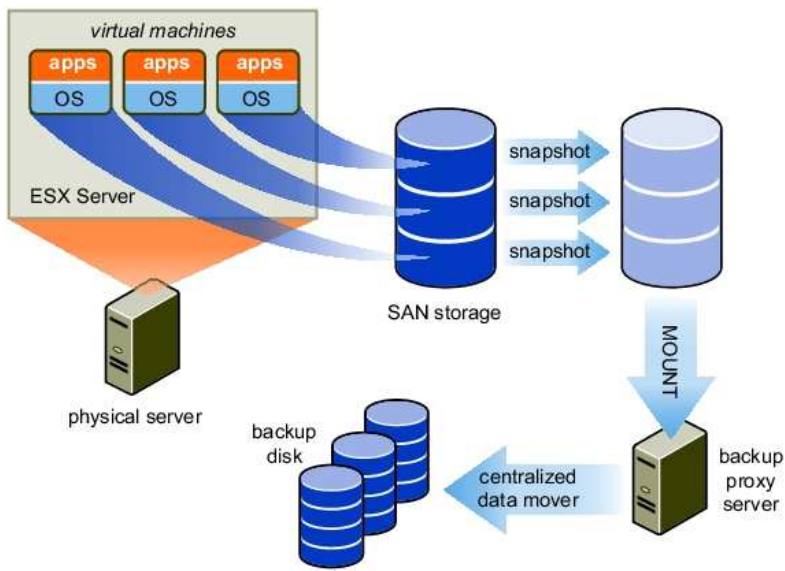

*Fig. 2.2 Backup infrastructure*

## 2.7 High availability systems

High availability solutions can be categorized into local and geographically distributed solutions.
Local solutions provide high availability in a single computing center.
Geographically distributed solutions, also known as disaster recovery solutions, are usually distributed deployments, able to protect application systems from natural disasters or network outages.

Amongst all possible kinds of failures, software and hardware (single node) failures as well as human errors can be protected by local high availability solutions. [2]
Local solutions will be the core of this dissertation. Local high availability solutions ensure a certain level of availability in a single computing center deployment.
To solve the high availability problem, a number of technologies and strategies are needed. The most used solution is based on redundancy, where high availability comes from redundant hardware systems and components. Local high availability solutions can be categorized into active-active solutions and active-passive solutions.

Fig. 2.3 [in2p3.fr] shows the two high availability dual server solutions: active-active vs active-passive.

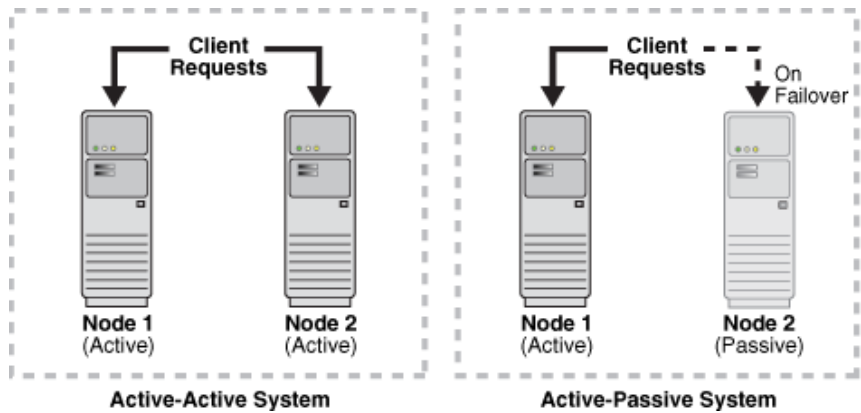

Fig. 2.3: High availability systems: active-active, active-passive

### 2.7.1 Active-Active

Active-Active solutions are characterized by the presence of two or more active nodes, able to balance the process load among them.
This solution deploys two or more active system instances. It can be used to improve scalability as well as provide high availability. All instances handle requests concurrently. In normal operation state, where none of the members have failed, all the cluster nodes are active and none is on standby state.
Traffic intended for the failed node is either passed onto an existing node or balanced across the remaining nodes. This is possible only if all the nodes utilize a homogeneous software configuration and are able to provide the same service.
Active-Active solutions are also generally referred to as cluster computing.

### 2.7.2 Active-Passive

Active-Passive, or Active-Standby, solutions are characterized by the presence of a primary node providing service, and a secondary node sleeping until it has to take over from the primary in case of damage of the primary node.
This solution deploys an active instance, responsible of handling requests, and a passive instance, that is on standby state. In addition, a heartbeat mechanism is set up between these two instances. This mechanism is provided and managed through clusterware applications.

The Active-Passive solution provides a fully redundant instance of each node. This configuration typically requires an amount of extra hardware. Consequently this solution entails a twofold increase of the hardware cost.

Usually cluster agents are also able to automatically monitor and enforce a failover solution between cluster nodes; this way, when the active instance fails, an agent shuts down the active instance completely, switching off the power supply too, and brings up the passive instance, so that applications can successfully resume processing.
A cluster agent is a software running on a node of the cluster, that coordinates availability and operations - such as load balance - with other nodes.
Active-Passive solutions are also generally referred to as heartbeat, or cold failover clusters.

## 2.8 Controller

A remote controller (or shared among all the nodes of a cluster) is needed to control the state of cluster nodes, processes and applications. Such controllers are also generally referred to as agents.
An agent is responsible for processing death detection and automatic restart, or actions aimed at the system recovery.
Processes may die unexpectedly, due to hardware breakdown, configuration or software problems. A proper process - the cluster agent – able to monitor and restart the service should constantly check all system processes and restart them if any problems appear.

## 2.9 State of the art

Since it is impossible to totally prevent the breakdowns of a computer system, one possible solution consists in setting up a redundancy mechanisms by duplicating critical resources, hardware, software. Cluster computing, in one of the several possible configurations, may accomplish these needs.
To implement a failover solution, at the moment several technologies are available. The most used are solutions such as Heartbeat (active-passive) and Cluster computing (active-active).

## 2.10 Linux-HA: Heartbeat

The Linux-HA (High-Availability Linux) project [3] provides a high-availability solution for Linux, FreeBSD, OpenBSD, Solaris and Mac OS X which promotes reliability, availability, and serviceability. The project's main software product is Heartbeat, a GPL-licensed portable cluster management program for high-availability.
Heartbeat can detect node failures reliably in less than half a second. With a low-latency communication infrastructure, such as Infiniband or Myrinet, this time could be lowered significantly.
The architecture is based on an active-passive high availability solution. Each service under high availability needs at least two identical servers: a primary host, on which the service run, one or more secondary hosts, able to recover the application in less than one second.

As a result of a failure detection, the active-passive roles are switched. The same procedure can be done manually, for planned or unplanned down time, i.e. in case of maintenance needs.

A heartbeat keep-alive system is used to monitor the health of the nodes in the cluster. Heartbeat monitors node health through communication media, usually serial and Ethernet links. It is a good solution to have multiple redundant connection links. Each node runs a heartbeat daemon process. When a node death is detected, Heartbeat runs a script to start or stop services on the secondary node.

A local disaster recovery solution is typically composed of two homogeneous nodes, one active and one passive. The active node is usually called master or production node, and the passive node is called secondary or standby node. During normal operation, the only working node is the master node; in the event of a node failover or switchover, the standby node takes over the production role, by taking its IP number, and completely replacing the master one.

To maintain the standby node for failover, the standby node contains homogeneous installations and applications: data and configurations must also be constantly synchronized with the master node.

Fig. 2.4 [linux-ha.org] shows a typical dual server heartbeat system.

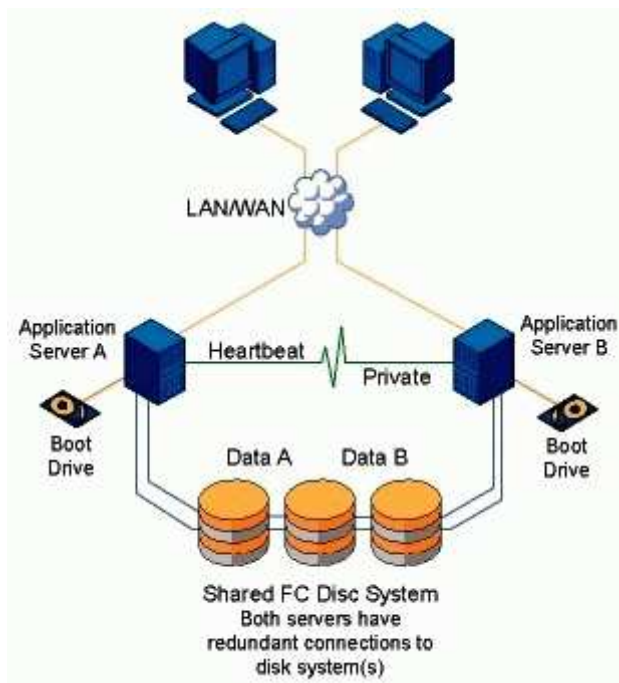

*Fig. 2.4: Heartbeat*

## 2.11 Cluster computing

Cluster computing is a system that allows its components to be viewed functionally as a single entity, from the point of view of a client for runtime services and manageability.
A cluster can be viewed as a set of processes running on multiple computers sharing the same environment. There is a very close correlation between clustering and redundancy. A cluster provides redundancy for a system and can be used for a high availability system [4].
High availability clusters (also known as HA Clusters or Failover Clusters) are computer clusters implemented to provide high availability of services. They operate by having redundant computers or nodes which are used to provide service when a system components fails.

When multiple instances of identical services are available, the client requests to these components can be balanced. This way ensures that all the application instances have approximately the same work load. With a load balancing mechanism running on site, all the instances are redundant. If any of the instances fails, the requests can be automatically sent to the surviving instances in the cluster. For this to work, there must be at least one extra component (a component in excess of the service's capacity).
This approach is less expensive and more flexible than failover approaches as Heartbeat, where a single live component is paired with a single backup component that takes over in the event of a failure. An analogous in a RAID disk controller technology, using RAID1 (mirror) is analogous to the "live/backup" approach to failover, whereas RAID5 is analogous to load balancing failover.
A cluster computing system may be used for several purposes, such as load balancing and failover.

Fig. 2.5 shows a Cluster computing architecture with a double redundant switch infrastructure.

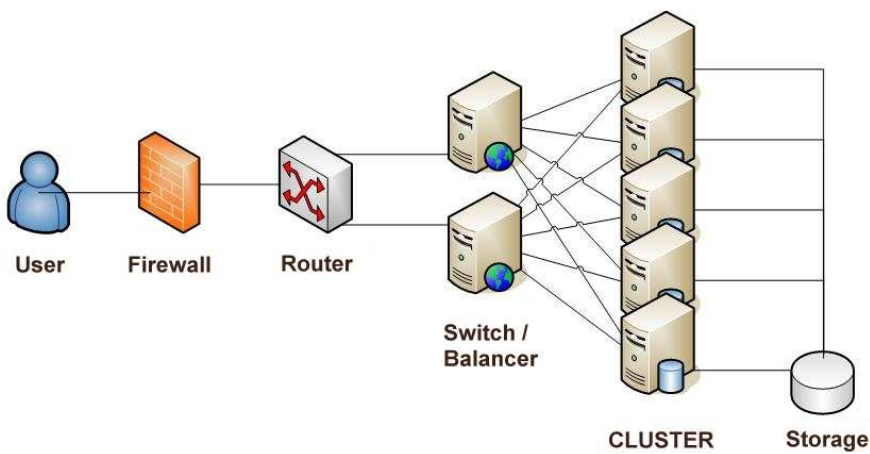

*Fig. 2.5: Cluster computing*

## 2.12 Load Balancing

In computer networking, load balancing is a technique to spread work between two or more computers, network links, CPUs, hard drives, or other resources, in order to get optimal resource utilization, maximize throughput, and minimize response time.
The use of multiple components and devices with load balancing technique, instead of a single component, may increase reliability through redundancy. The balancing service is usually provided by a dedicated program or hardware device (such as a multilayer switch).

Fig. 2.6 [primustel.ca] shows a load balanced architecture.

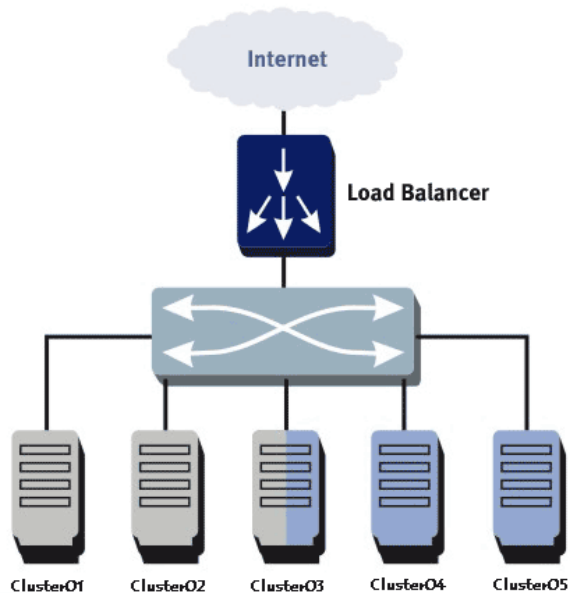

*Fig. 2.6: Load balance*

## 2.13 Failover

Failover is the capability of a complex system to switch over automatically to a redundant or standby computer server, system, or network upon the failure or abnormal termination of the previously active server, system, or network. Failover must happen in a completely automatic manner, without human intervention.
Load balancing is often used to implement failover. All the system components are continuously monitored (e.g. a server may be monitored by ping, a web service by fetching a known page), and when one component becomes non-responsive, the main controller is informed and no longer sends traffic to it.

After hardware/software fault detection, high availability failover cluster remedies this situation immediately by restarting the application on another system, without requiring human intervention. In case the broken component comes back on line, the load balancer can decide to reintroduce it again in the cluster or not, and if necessary begins to route traffic to it again.

A high availability cluster implementation attempts to build redundancy into a cluster, in order to eliminate single points of failure, including multiple network connections and data storage redundancy [5].
Another possible solution to implement a failover architecture is to devolve the recover task upon a switch able to assign the same IP address to a secondary server, in case of primary server failure.

Fig. 2.7 shows a Failover solution managed via switch (layer 2).

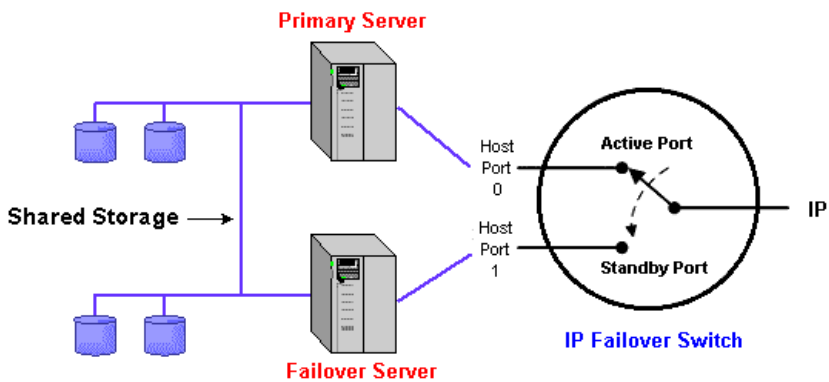

*Fig. 2.7: Failover*

## 2.14 Related issues: Split-brain

Every clustering software must be able to handle the "split-brain" problem. Split-brain occurs when all the private links go down simultaneously, but the cluster nodes are still running. If it happens, each node in the cluster may mistakenly decide that every other node has gone down and therefore may attempt to start services that other nodes are still running. Having duplicate instances of services may cause data corruption on the computing model or shared storage.
To prevent this impasse, every time a controller switches on a secondary service or host in substitution of the main one crashed, it attends at the same time to definitely kill the main service; this is usually done by switching off the power supply of the server owning the interested service.
High availability clusters usually use a heartbeat private network connection to monitor the health and status of each node in the cluster.

## 2.15 Application design requirements

Not every application can run in a high-availability cluster environment. The necessary architecture decisions need to be made in the first software design phase.
In order to run in a high-availability cluster environment, an application must satisfy at least the following technical requirements:
- easy way to start, stop, force-stop, status check - usually via command line interface or script, including support for multiple instances of the application;
- use of shared storage (Network Attached Storage NAS or Storage Area Network SAN);
- store as much as possible of its state on non volatile shared storage - it is needed to restart the application on another node at the last safe state before failure, using the saved state from the shared storage;
- avoid data corruption after a crash.

## 2.16 Computing center reliability

High availability clusters usually utilize all available techniques and architectures to make systems and infrastructure as reliable/available as possible. These include:
- disk mirroring: RAID 0, 1, 5, 6 depending on the application safety requirements and users needs;
- redundant network connections with redundant components, so that single cable, switch, or network interface failures do not entail network outages;
- storage area network (SAN) with redundant data path and data connections, to avoid loss of connectivity to the storage;
- redundant electrical power supply on different circuits;
- uninterruptible power supply (UPS).

All these features together help to minimize the chances of a clustering failover.

## 2.17 Best practices

In addition to architectural redundancies, the following local high availability technologies are also necessary in a comprehensive high availability system:
- to reduce the number of components or devices apt to break, it is useful to remove all unnecessary peripherals, such as keyboard, mouse, graphics and audio card, CD, USB devices;
- a thermal welding between the network interfaces of the nodes involved in a high availability cluster system is an additional safety against possible disconnections or not well fixed links.

## 2.18 Addenda

A cluster agent should also store the number of restarts within a specified time interval. This is very important, since a large number of restarts in a short time may lead to additional faults or failures. Therefore a maximum number of restarts within a specified time interval should also be planned.
A cluster of similar nodes often needs to share a common configuration. A proper configuration management ensures that components provide the same reply to the same incoming request, allows these components to synchronize their configurations, and provides highly available configuration management for less administration downtime.

# 3 VIRTUALIZATION

## 3.1 Virtualization definition

The term virtualization was coined in the 1960s, to refer to a virtual machine [6]. The creation and management of virtual machines has been called platform or server virtualization [7].
A virtualization system is an architecture able to separate an operating system from the underlying platform resources.
Server virtualization is performed on a given hardware platform by introducing a software layer, which creates a simulated computer environment, a virtual machine, for its guest software. The guest software runs just as if it were installed on a stand-alone hardware platform. The guest software itself is in many cases a complete operating system, and can also be different from the operating system hosted by the physical machine.
Usually, many virtual machines run on a single physical machine; their number is limited by the host hardware capability, such as core number, CPU power, RAM resources. There is no requirement for a guest operating system to be the same as the physical host one.
The guest system is able to access specific peripheral devices, exploiting interfaces to those devices, e.g. hard disk drive, network interface card, graphic and audio card.

Fig. 3.1 [VMware] shows the four level of a virtualized architecture - from the bottom to the top: hardware and operating system of the host machine, virtualization layer, operating system and applications of the virtual hosted machines.

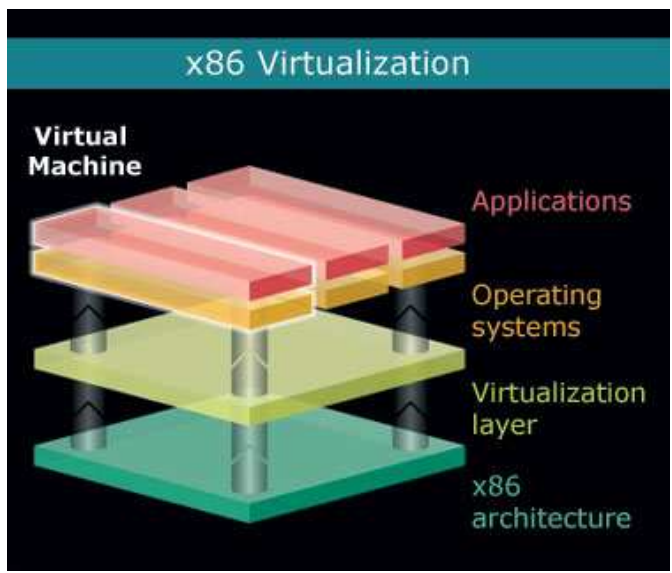

*Fig. 3.1: x86 architecture layers*

## 3.2 Rings and protection domains

In computer science, hierarchical protection domains, also called protection rings, are a mechanism to protect data and functionality from faults and malicious behavior.

Computer operating systems provide different levels of access to resources. A protection ring is one of the hierarchical levels of privilege within the architecture of a computer system. This is generally hardware enforced by some CPU architectures providing different CPU modes at the firmware level.

Rings are arranged in a hierarchy from most privileged (most trusted, the ring 0) to least privileged (least trusted, the ring 3). Ring 0 is the level with the most privileges and interacts directly with the physical hardware such as CPU, memory, devices. The kernel of an operating system runs in ring 0.

### 3.2.1 The ring problem

The virtualization software, to be able to directly access the hardware, could run in ring 0. The problem is how to run privileged operations in operating systems hosted in virtual machines [8].

Fig. 3.2 [Wikipedia] shows the four operating system ring levels.

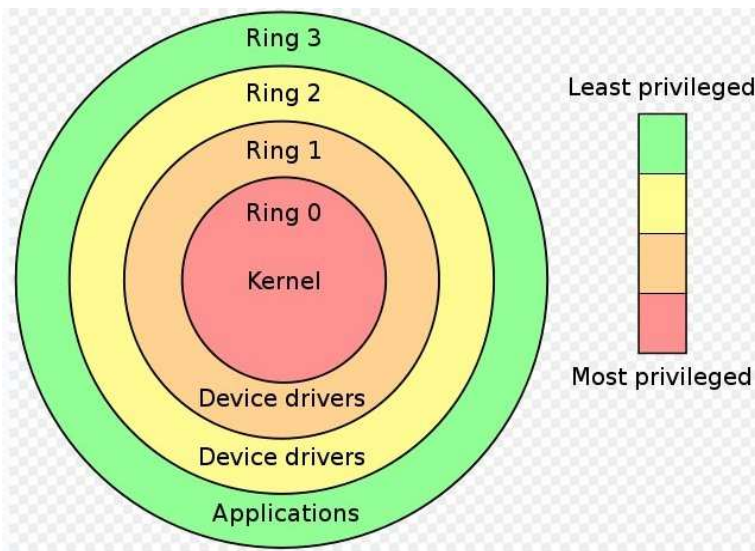

Fig. 3.2: x86 ring levels

## 3.3 Virtualization solutions

A virtualization architecture provides a CPU virtualization system as well as memory, device and I/O virtualization component. This involves sharing the physical system memory and devices, and dynamically allocating them to virtual machines.

Several virtualization solutions are available today, meeting the requirements of several needs.

Fig. 3.3 [zdnet.com] shows the differences between a non-virtualized and a virtualized computer architecture.

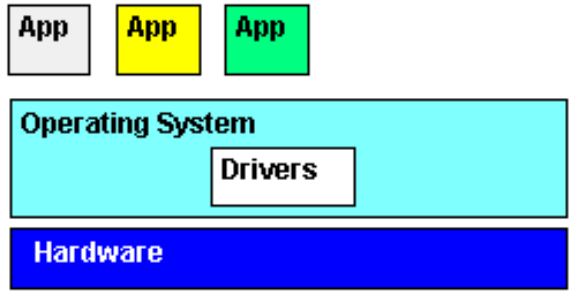

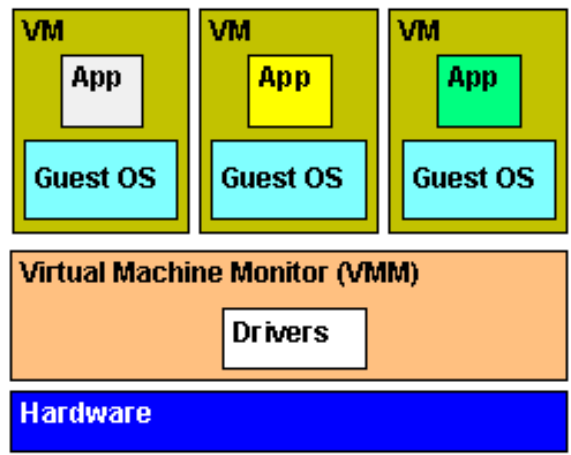

*Fig. 3.3: Virtualization architecture*

## 3.4 Full virtualization

In full virtualization, the virtual machine simulates enough hardware to allow an unmodified guest operating system to be run in isolation. This approach was pioneered in 1966 with the IBM mainframes.
Full virtualization fully abstracts the guest operating system from the underlying hardware (completely decoupled). The guest operating system is not aware it is being virtualized and requires no modification.
Full virtualization offers the best isolation and security for virtual machines, and allows simple procedures for migration and portability as the same guest operating system instance can run virtualized or on native hardware.

While the virtualization layer is executed at the application level, the ring problem is solved by the full virtualization systems with the Binary Rewriting technique: the flow of the binary instructions performed by the virtual machines is inspected by the hypervisor and the privileged instructions are conveniently translated. The Binary Rewriting technique is computationally highly expensive.

More recent full virtualization products include Parallels [9], VirtualBox [10], Virtual PC [11], Virtual Server [12], Hyper-V [13], VMware [14], QEMU [15].

Fig. 3.4 [Tosslab.it] shows a Fully virtualized architecture.

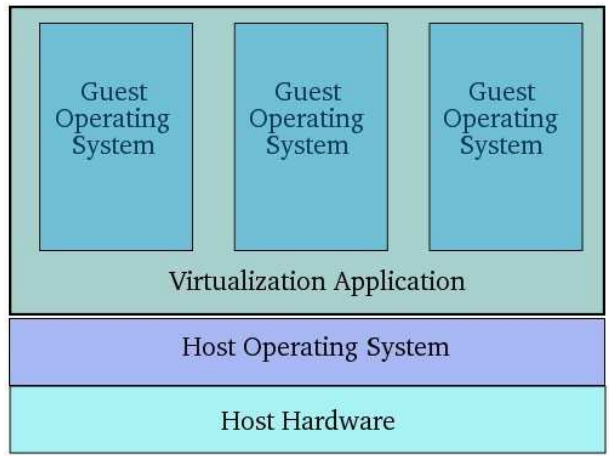

*Fig. 3.4: Full virtualization*

## 3.5 Paravirtualization

In a paravirtualized architecture, a light software layer (Hypervisor) runs directly over the hardware. The Hypervisor is able to allocate the resources needed by the virtual machines. A privileged operating system instance runs over the Hypervisor in order to manage all the active virtual machines.

Paravirtualization enables multiple isolated and secure virtualized servers running over the same physical host.

A paravirtualized system provides a low virtualization overhead, but the performance advantage of paravirtualization over full virtualization can vary greatly depending on the workload.

The host and guest operating systems have to be modified in order to replace the privileged operations with Hypervisor calls (hypercalls). While the Hypervisor is executed at ring level 0, the guest operating systems are executed at level 1.

As paravirtualization cannot support unmodified operating systems (e.g. Windows), its compatibility and portability are limited.

Examples of paravirtualization systems are XEN [16] and VMware Infrastructure [14].

Fig. 3.5 [Tosslab.it] shows a Paravirtualized architecture.

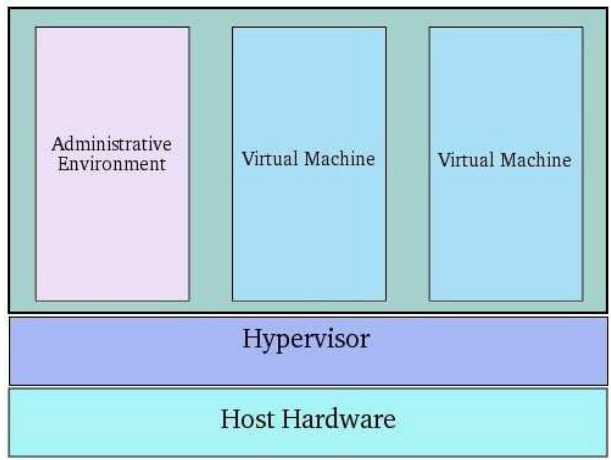

*Fig. 3.5: Paravirtualization*

A XEN system is structured with the Xen Hypervisor as the lowest and most privileged layer. Above this layer are one or more guest operating systems, which the hypervisor schedules across the physical CPUs. The first guest operating system, called "domain 0" (dom0), is booted automatically when the hypervisor boots and is given special management privileges and direct access to the physical hardware. The system administrator can log into dom0 in order to manage any other guest operating systems, called "domain U" (domU) in Xen terminology. The device accesses from all the domU are managed through the dom0.

Fig. 3.6 [zdnet.com] shows the dom0 - domU architecture in a paravirtualized system.

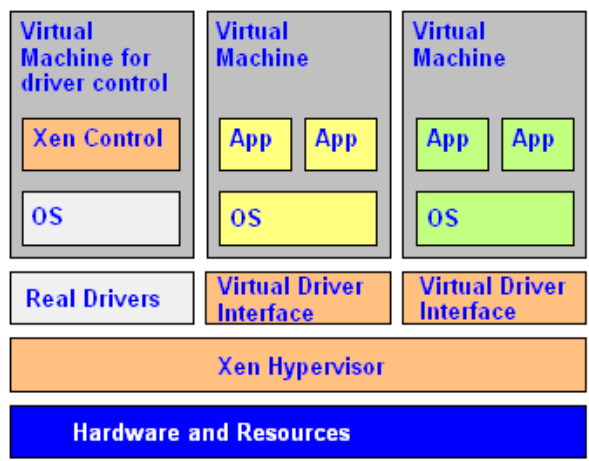

*Fig. 3.6: Paravirtualization dom0 - domU*

## 3.6 Shared kernel - Operating system level virtualization

Shared kernel virtualization, also known as Operating system level virtualization, is a virtualization method where the kernel of an operating system allows for multiple isolated user-space instances, instead of just one. Such instances, called containers, may look like a real server, from the point of view of the end users.
In addition to isolation mechanisms, the kernel is able to provide resource management features to limit the impact of one container's activities on the other containers. This way multiple instances can run exactly as multiple physical servers.
The guest (virtual machine) operating system environments share the same operating system as the host system (physical machine). To be more precise the same operating system kernel is used to implement the guest environments. Applications running in a given guest environment view it as a stand-alone system.
The uniqueness of the running kernel solves the ring problem, but forces all the guest operating systems to be compatible with the host operating system.
Examples of Operating system level virtualization systems are Parallels Virtuozzo Containers [17], OpenVZ [18], Solaris Containers [19], FreeBSD Jails [20].

Fig. 3.7 [Tosslab.it] shows a Shared kernel - Operating system level virtualization architecture.

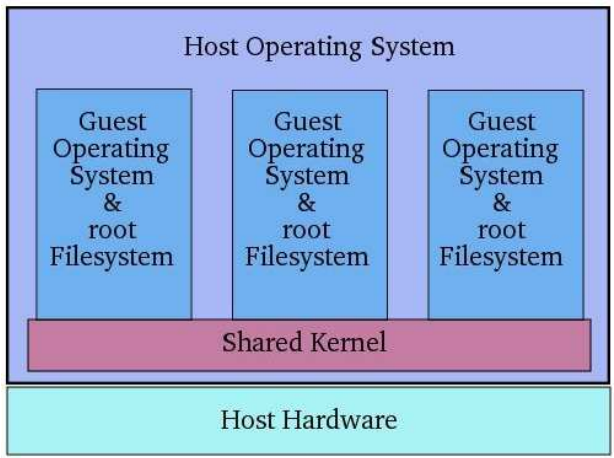

*Fig. 3.7: Shared kernel - Operating system level virtualization*

## 3.7 Hardware-assisted virtualization

In hardware-assisted virtualization, the hardware provides architectural support (libraries included in hardware) that facilitates building a virtual machine monitor and allows guest operating systems to be run in isolation.
In 2005 Intel and AMD provided additional hardware to support virtualization: AMD-V [21] virtualization architecture for AMD, VT-x [22] virtualization architecture for Intel.
The implementations of hardware-assisted virtualization are: VMware Workstation, Xen 3.x, Linux KVM and Microsoft Hyper-V. Hardware assisted virtualization solutions allow to guest unmodified operating systems, by introducing a new ring level (-1) in the ring stack running the Hypervisor - the kernel may be executed at ring level 0.

Fig. 3.8 [Tosslab.it] shows a Hardware-assisted architecture.

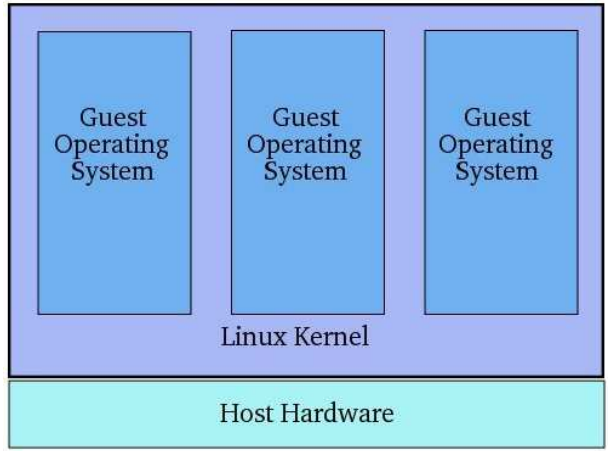

Fig. 3.8: Hardware-assisted virtualization

## 3.8 Virtualization features

### 3.8.1 Server consolidation

Even if a computational node may need an ever increasing number of processors, most of the times a service needs at most only one processor.
Virtualization can meet this demand, by decoupling software from hardware and splitting a multi processor server into more independent virtual hosts, for a better utilization of the hardware resources, allowing services to be distributed one per processor.
In case of server consolidation, many small physical servers are replaced by one larger physical server, to increase the utilization of expensive hardware resources. Each operating system running on a physical server is converted to a distinct operating system running inside a virtual machine. This way a single large physical server can host many guest virtual machines.

Fig. 3.9 shows a typical example of server consolidation in a computing center.

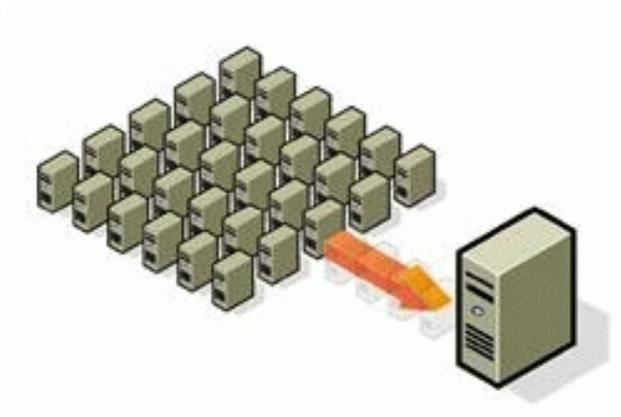

Fig. 3.9: Server consolidation

### 3.8.2 Management ease

A virtual machine can be more easily managed, configured and controlled from outside than a physical one. While it is possible to switch off a physical host remotely, it is impossible to switch on a turned off physical machine. In virtual environment, you can manage all the actions on virtual machine from a remote location, including the power on and power off. This is very useful to test a new kernel or for teaching courses.

## *3.9 Virtual machine migration*

A virtual machine can easily be moved from one physical machine and relocated to another, if needed [23]. Typical examples of virtual machines movement are: virtual machine cloning (to copy more instances of the same virtual machine), virtual machine relocation (to free the original machine hosting the virtual machine), in case of maintenance or break [24].
Because of their mobility, virtual machines can be used in high availability and disaster recovery scenarios.

Fig. 3.10 [VMware] shows a virtual machine movement - migration from one physical server to another physical server.

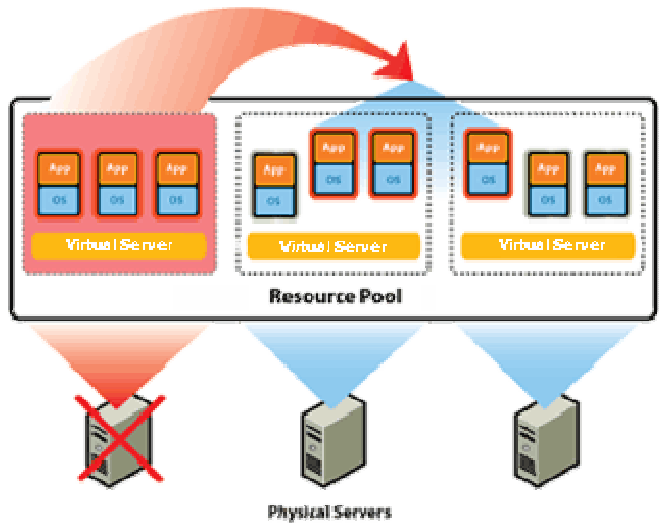

*Fig. 3.10: Virtual machine migration*

## 3.10 Comparison of available products

### 3.10.1 XEN

Xen [25] is a virtual machine monitor for IA-32, x86, x86-64, IA-64 and PowerPC 970 architectures. It allows several guest operating systems (including Windows, Linux, Solaris, and various versions of the BSD) to be executed on the same computer hardware concurrently. Xen provides paravirtualization.
Xen was initially created by the University of Cambridge Computer Laboratory and is now developed and maintained by the Xen community as free software, licensed under the GNU General Public License (GPL2).

With Xen virtualization, a thin software (the Xen hypervisor) is inserted between the server hardware and the operating system. The hypervisor provides an abstraction layer that allows each physical server to run one or more virtual servers, decoupling the operating system and its applications from the underlying physical server [26].

### 3.10.2 VMware

Vmware [27] is an enterprise product. It uses two different virtualization approaches, using a hosted or a hypervisor architecture. Vmware provides both full virtualization and paravirtualization.
A hosted architecture installs and runs the virtualization layer as an application on top of the operating system. It guarantees a great flexibility level.

A hypervisor architecture (bare-metal) installs the virtualization layer directly on a virgin x86-based system. Since a hypervisor architecture has a direct access to the hardware resources rather than going through an operating system, it is more efficient than a hosted architecture, and it is able to deliver greater performance. Examples of hosted architecture are VMware Player and VMware Server. Example of hypervisor architecture is VMware ESX Server.

# 4 SCENARIO: GRID DATA CENTER

## 4.1 CERN Large Hadron Collider - Geneva

The Large Hadron Collider (LHC) is the world's largest and highest-energy particle accelerator, intended to collide opposing particle beams, protons at an energy of 7 TeV/particle or lead nuclei at 574 TeV/particle.
The Large Hadron Collider was built by the European Organization for Nuclear Research (CERN) with the intention of testing various predictions of high-energy physics, including the existence of the hypothesized Higgs boson and of the large family of new particles predicted by supersymmetry. It lies in a tunnel 27 kilometers in circumference, as much as 175 meters beneath the Franco-Swiss border near Geneva, Switzerland. It is funded by and built in collaboration with over 10.000 scientists and engineers from over 100 countries as well as hundreds of universities and laboratories.
The Italian National Institute of Nuclear Physics INFN and Scuola Normale in Pisa participate in the LHC project, with respect to research activity in physics, engineering and information technology.

## 4.2 LHC Computing model

Several Petabytes of data will be collected per year by each of the LHC experiments. The computing power to process these data, and to produce and process the comparable amounts of simulated data required for analysis, is estimated to be equivalent to something like 100.000 of today's personal computers.
These strong hardware requirements lead naturally to the idea of a computing model based on Grid [28] [29], whereby geographically dispersed computing resources are shared together.

CERN will not have the resources to provide all LHC computing needs. LHC experiments will therefore make use of a computing infrastructure which is intended to allow transparent access to data sets and computing resources regardless of the location. In this respect the correlation with the Grid project [30] [31] is clearly very strong. Grid-like applications are such applications that are run over the wide area network using several computing and data resources where each of the single computing resources can itself be a network of computers.

The purpose of the LHC Computing Grid (LCG) is to provide the computing resources needed to process and analyze the data gathered by the LHC Experiments.
The LCG project, aided by the experiments themselves, is addressing this by assembling at multiple inter-networked computer centers the main offline data storage and computing resources needed by the experiments and operating these resources in a shared Grid like manner. One of the project's most important goals is to provide common software for this task and to implement uniform means of accessing resources. It is useful to classify the computer centers functionally in "Tiers" [32] [33] [34].

Fig. 4.1 [INFN Pisa] shows the CERN - EGEE Tiers computing model.

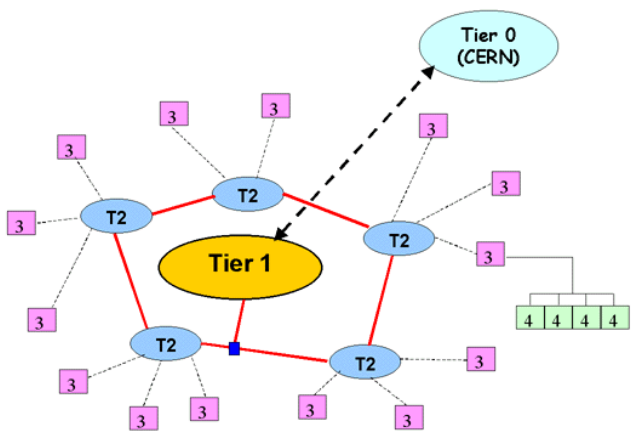

*Fig. 4.1: Hierarchy of resources: Tier0, Tier1, Tier2, Tier3, Tier4*

## *4.3 Tier0*

Tier0 is at CERN. It receives the raw and other data from the experiments online computing farms and records them on permanent mass storage. It also performs a first pass reconstruction of the data. It will hold a complete archive of all raw data. The raw and reconstructed data are distributed to the Tier1 Centers.

### 4.3.1 Tier1

Tier1 Centers provide a distributed permanent backup of the raw data, permanent storage and management of data needed during the analysis process, and offer a Grid-enabled data service. They also perform data-heavy analysis and re-processing, and may undertake national or regional support tasks, as well as contribute to Grid Operations Services. A Tier1 is a regional computing center; at least one Tier1 center is foreseen for the major countries involved (e.g. USA, Italy, France, UK).

### 4.3.2 Tier2

Tier2 centers provide well managed, Grid enabled disk storage and concentrate on tasks such as simulation, end-user analysis and high performance parallel analysis. It is a regional center similar to a Tier1 center but on a smaller scale. Its services would be more focused on local data analysis requirements, and on distributed Monte Carlo production.

### 4.3.3 Tier3

A Tier3 center is usually comprised of an Institute or University computing center.

### 4.3.4 Tier 4

A Tier4 is an individual desktop.

## 4.4 Tiers Transfer rate

The Tiers model approach developed at CERN is based on a robust and reliable network infrastructure, able to move data at a sustained rate of some Gb/s average.

The CERN Tier 0 computing center process several thousand concurrent batch jobs simultaneously and exports the raw and processed data to 10 Tier 1 sites at about 10G/s rate. When all experiments will take data together, the aggregate rate out of CERN will be 50 Gb/s and about 40 TB of data will be stored each day.

Fig. 4.2 [CERN] shows the Tiers architecture and the expected transfer rate between each couple of levels.

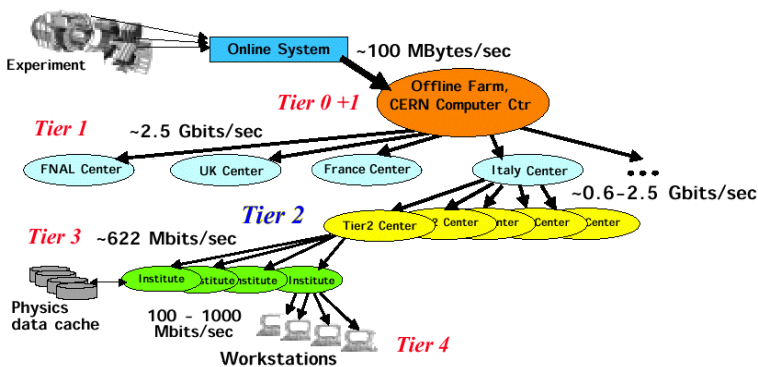

*Fig. 4.2: Tiers architecture and transfer rate*

## 4.5 Grid data center

INFN-PISA participates in the LHC experiments as a Tier2 regional center. The computer center [35] consists of 2.000 CPU, 500 TB disk and a full 1 GB switching infrastructure, and it is expected to reach more than 5.000 CPUs and 1 PB of disk space in the next two years.

We decided to implement the Tier 2 as part of a common infrastructure between INFN, Scuola Normale, and Physics Department of Pisa University [36]. By using a suitable job fair sharing among all the contributing groups, it has been possible to allocate all the computational power under the Grid control.

The advantages of such an approach are:
- the power off of all the individual farms shared all around the department;
- the management ease and centralization, with a great save in human time for installation, upgrade, administration.

## 4.6 Architecture

Distributed or Grid computing in general is a special type of parallel computing that relies on complete computers (with on board CPU, storage, power supply, network interface, etc.) connected to a network by a conventional network card interface, such as Ethernet, without any particular sophisticated device. This is in contrast to the traditional notion of a supercomputer, where many processors are connected together by a local high-speed computer bus, or parallel computing, where the processors are connected by a low latency network, such as Infiniband or Myrinet.

The scalability of geographically dispersed Grids is usually guaranteed by the low need for connectivity between nodes with respect to the capacity of the public Internet.
The middleware provided by LCG [37] / EGEE [38] is based on the Globus project [39] [40] [41]. Globus is a community of organizations and individuals developing fundamental technologies behind the Grid.

A Grid computing center usually consists of:

- **Computing Element** - CE: it is the back-end of the system. It acts as an interface between the user and the Grid environment. The Computing Element in a Grid computing center is also a batch queue balancer to a centrally managed farm of computers.

- **Storage Element** - SE: provides uniform access to storage resources. It could be simply a disk server, large disk arrays or Mass Storage System such as dCache or Castor. The data access authorisation is handled using Grid credentials (certificates). File transfer is managed through Grid Security Infrastructure File Transfer Protocol GSIFTP, while the native Storage Resource Manager SRM protocol allows POSIX-like data access.

- **Worker Node** - WN: the computational node, able to run Grid jobs.

- **User Interface** - UI: a computer which has the set of user level (typically Linux command line) client tools and API libraries installed on it to access the Grid environment. It is a gateway between Grid and the end-users. A user must have an account on the machine and her/his certificate installed.

- **Information System** - IS: a scalable and reliable information system able to publish all the Grid resources of a single site or a larger geographical region. The Globus Project use a Grid information system based on the LDAP system.

- **Resource Broker** - RB / Workload Management System - WMS: examines Requirements and Rank expressions (and also any data requirements) in the Grid job JDL. All CE are filtered against the Requirements, and the Rank is calculated for all the CE which match; on the basis of Rank it determines which CE to submit the job to [42]. For that purpose, the RB / WMS must retrieve information from the IS and the File Catalog.

- **LCG File Catalog** - LFC: a service providing mappings between Logical File Names Grid Unique IDentifiers and Storage URLs.

- **Replica Catalog** - RC: contains the information about stored and replicated data.

- **Virtual Organisation Membership Service** - VOMS: provides administration services for users and administrators of a VO. VOMS is essentially an authentication service: the list of VO users authorized to use VO resources comes from the VOMS and is propagated to the resources (RB, CE, SE, etc).

Fig. 4.3 [CERN LCG] shows the architecture of the LCG Grid infrastructure.

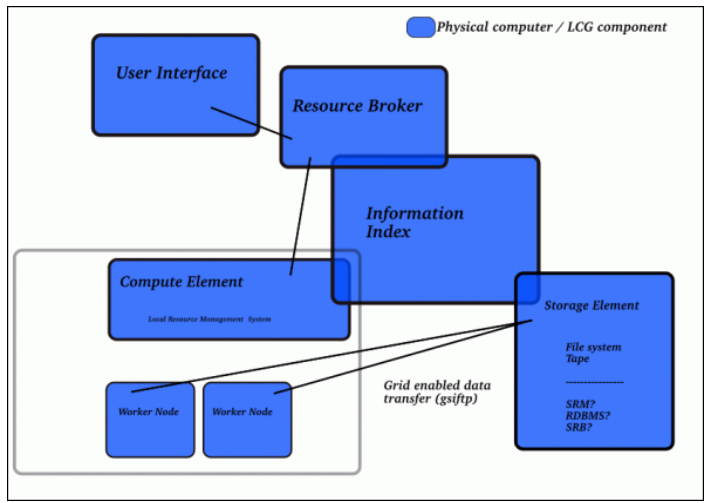

*Fig. 4.3: Grid elements*

A Grid environment is based also on:

- **Virtual Organization** - VO: the base idea of a Grid environment. A Virtual Organization is a group, typically an experiment, who shares the same computing resources. A VO software manager can install centrally the source code to make it available to all the nodes.

- **Job Definition Language** - JDL: is the language used to specify the resources required by a job. To submit a job to the Grid a JDL file is created and passed to a RB (Resource Broker) or a WMS (Workload Management System) that examines the JDL and determines, with the help of the IS (Information Service), the best CE (Computing Element) to run the job.

Fig. 4.4 [CERN LCG] shows the flowchart of a Grid Job submission.

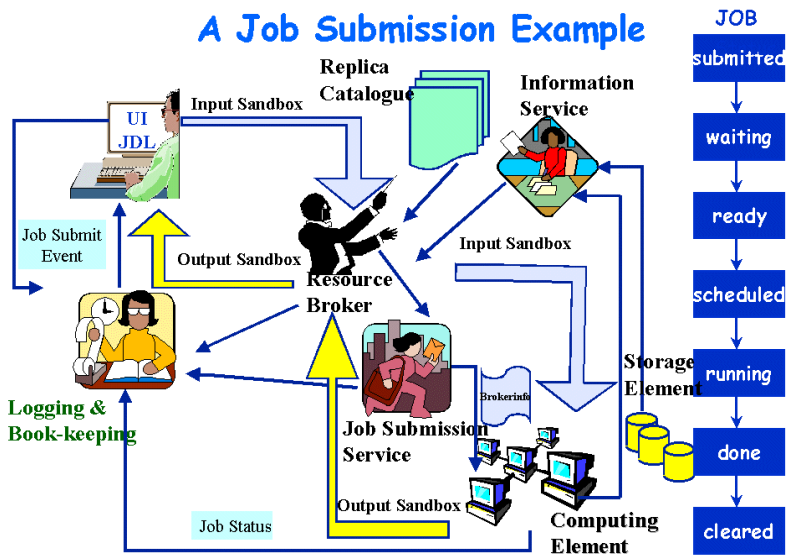

*Fig. 4.4: Grid Job submission*

- **GridFTP**: the standard protocol for Grid based file transfers.

## *4.7 Relaxed high availability model*

The primary idea on which is based the Grid infrastructure is that each node can be purchased as a commodity hardware. A Grid computing center is therefore comprised of low cost devices. Based on the idea that each central service is critical for the single computing center (not for the whole Grid), but it can bear a system down of 5 to 10 minutes, I introduced the new concept of "relaxed" high availability.
A new approach to guarantee a relaxed high availability level to all the services will be shown in the next chapters.

# 5 INFRASTRUCTURE

A complex infrastructure has been deployed in the last years in order to improve the computing center management architecture. The aim of such an infrastructure is to ease and centralize the complete computing center maintenance, from hosts installation to virtual machine upgrade [44].

## 5.1 PXE architecture

The Preboot eXecution Environment (PXE) [43] is an environment to boot computers using a network interface, independently of available data storage devices (e.g. hard disks) or installed operating systems.
PXE is an open industry standard developed by a number of software and hardware vendors (Intel, 3Com, HP, Dell, Compaq, and others).
PXE works with a network interface card NIC in the target host, and transforms the NIC into a boot device. The PXE vision is to "Make the network interface a standard, industry-accepted PC boot device".
By adding the NIC to the list of traditional standard boot devices - such as floppy drives, pen drives, hard disks, and CD-ROMs - a new host is able to load the operating system or set up programs directly from network. It allows the host client to do a "network boot".
PXE boots the client host from the network by transferring a boot image file from a server. This file can be a minimal operating system for the host or a preboot chain loader linked to a boot in a preexisting disk operating system.
Because PXE works with the network interface card NIC, it requires a PXE enabled NIC. The PXE protocol is supported by almost all the network interfaces available today off the shelf.

## 5.2 System requirements

In order to implement an architecture able to bootstrap a new host via network PXE, there are no other requirements than servers usually involved in the daily practice of a computing center, such as DHCP, TFTP, and HTTP servers.
By modifying the DHCP configuration file, a string is passed to the host at the boot moment.

## 5.3 How does PXE work?

The PXE boot process extends the Dynamic Host Configuration Protocol DHCP with the additional information needed to a computer remote boot. This information includes the client vendor and class enabling the PXE server to select a client-specific image.
The system requesting a PXE boot uses the DHCP DISCOVER message to identify its vendor and machine class and to request the location and the name of an image file. The PXE client identifies its vendor and class of machine because there might be multiple images available through the PXE servers.

A PXE boot process involves many information exchanges:

1. The PXE client sends a DHCP DISCOVER with the PXE options filled in.

2. The DHCP server responds with a DHCP OFFER containing TCP/IP parameters.
3. The PXE client replies with a DHCP REQUEST.
4. The DHCP server responds with a DHCP ACK.
5. If the DHCP server is also the PXE server, the DHCP ACK will usually contain the TFTP server name and boot file name. If the PXE server is a different system, there is a separate exchange of requests and replies between the PXE server and the PXE client following the initial DHCP process.

Fig. 5.1 [esrf.eu] shows the PXE operations.

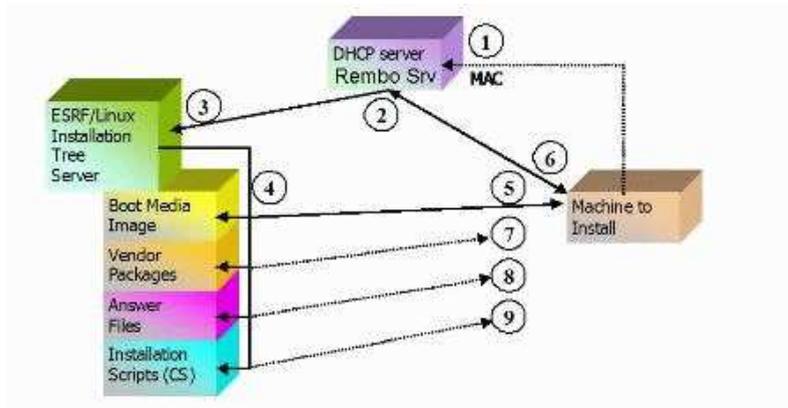

*Fig. 5.1: PXE operations*

The firmware on the client tries to locate a PXE service on the network, in order to receive information about available PXE boot servers. The PXE server watches for DHCP discovery requests including a special tag identifying the host as a PXE client. When the DHCP request includes the tag, the PXE server replies to the client with the configuration information, including the name and the address of a boot image file. The boot image file is transferred into the client RAM using the Trivial File Transfer Protocol TFTP - a very simple transfer protocol without any type of encryption - and this file is used to boot the host.

The PXE may be approximately seen as a combination of DHCP and TFTP, with small modifications to both of them. To start a PXE boot session, the PXE firmware broadcasts a DHCPDISCOVER packet extended with PXE specific options (extended DHCPDISCOVER) to port 67/UDP (DHCP server port).

Fig. 5.2 [os3.nl] shows the PXE information flow.

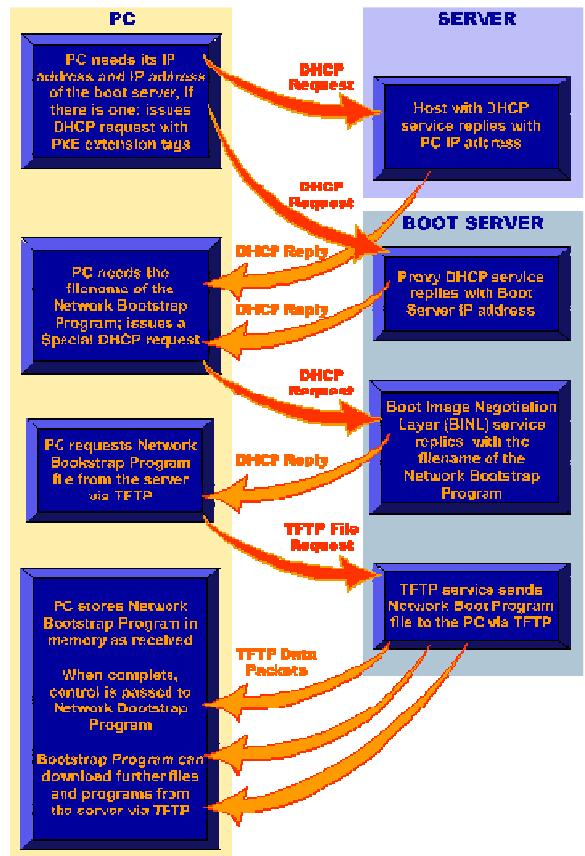

*Fig. 5.2: PXE information flow*

## 5.4 Advantage of network boot

There are many advantages in using a boot via network, such as:
- booting diskless systems - automating system maintenance such as backups, or system checking such as virus scanning;
- deploying software and operating systems for new systems.

In my research work, the network boot is intended to be used for installing an operating system in a brand new host with no operating system on board - or reinstalling the operating system in a host where the operating system has failed.

In order to install a new host, the only operation required is to connect the host to the network and power it on. This can dramatically reduce the administration time required for a new installation.

## 5.5 Auto-installer system

Exploiting the PXE architecture, at the Italian National Institute for Nuclear Physics and Scuola Normale Superiore - Pisa, a complete system has been developed to automatically install operating system and middleware. The system has been successfully used for the installation of 2000 CPUs progressing in parallel.

Due to a continuous development of a production environment, where scripts, programs and setup change almost every day, the chosen solution to have an up to date system, has been the implementation of an architecture where the installation of a host is performed by a single script.

The centralization of all the installation procedures offers an essential aid to the site administrator when the operations involve more than some dozens of hosts simultaneously.

## 5.6 An improvement to classical architecture

Exploiting an almost unused field in DNS record, the HINFO (Host Info) field, in INFN Pisa we have set up a post install script able to perform all the needed operations to install middleware, patches and applications on all the computing center servers.

In the HINFO field, a non compulsory field in DNS record, an indication is stored of what the machine is and what it runs. Once the operating system is installed via PXE architecture, the post install script reads the HINFO from the DNS server and install the needed middleware, coherently with the machine type indicated in HINFO note. In this way we completely automated not only the installation of the operating system, but also the installation of middleware, applications, and their complete configuration.

## 5.7 Update management

In order to have a large set of hosts up to date, an update management system has been designed at INFN Pisa section. It may happen that when an upgrade is needed, a set of hosts is down for maintenance, or broken. To avoid a corrupted state of those hosts at the start up moment - in a few minutes or after a month - an update management system is required.

A central repository contains all the scripts necessary to update all the cluster hosts, classified by host type (computational node, storage element, etc), and the date to be applied. When a host is switched on, it checks the central scripts repository and matches it with its internal update history, applying all the update scripts not yet performed.

The same system may be used also for a scheduled update of all the hosts in a computing center, by applying the same script over all the hosts at a given time.

## 5.8 A dual boot architecture via PXE

In order to achieve a dual boot system on a whole 1000 CPU cluster without the need of writing a single byte on the existing hard disk devices - it could be a constraint imposed by the owner of the cluster you are using - a new procedure has been developed at INFN Pisa, from an idea by Enrico Mazzoni.

Exploiting the PXE architecture and infrastructure used to install all the computing center hosts, and adding a single hard disk per host, we pass to the host the correct chainloader via PXE at the boot moment, directing the boot from one disk rather than the other.

To reach this goal, a MEMDISK file is given to the host at the boot time, to allow booting legacy operating systems (such as DOS) via PXE. It is used in conjunction with PXELINUX.

MEMDISK is part of the SYSLINUX suite, and provides support for booting legacy operating systems. The SYSLINUX Project covers lightweight bootloaders for MS-DOS FAT filesystems (SYSLINUX), network booting (PXELINUX), bootable El Torito CD-ROMs (ISOLINUX), and Linux ext2/ext3 filesystems (EXTLINUX).

MEMDISK simulates a disk by claiming a chunk of high memory for the disk and a (very small - 2K typical) chunk of low (DOS) memory for the driver itself, then hooking the INT 13h (disk driver) and INT 15h (memory query) BIOS interrupts. MEMDISK is a kernel module used in conjunction with one of the SYSLINUX bootloaders, in our case PXELINUX. There is the need of a disk image as well as the MEMDISK file itself. As far as the bootloader is concerned, MEMDISK is the kernel and the disk image is the initial ramdisk (initrd).

Using GRUB - a boot loader package from the GNU Project which allows a user to have several different operating systems on his computer at once, and to choose which one to run when the computer starts - the pxelinux.cfg file is used to select from different kernel images available on a particular disk device partitions.

One boot loader can boot another boot loader by chain loading. Chain loading is a method used by computer programs to replace the currently executing program with a new program, using a common data area to pass information from the current program to the new program. At this stage GRUB can also pass control of the boot process to another loader.

To select the boot operating system from one rather then the other disk, it is only needed to change the PXE link or the MAC address to IP association in DHCP configuration file, and reboot the host.

**pxelinux.cfg/disk1[2]**
```
  kernel kernel/memdisk
  append initrd=grub/grub-disk-hd1[2].img
```

**grub-disk-hd1[2].img/grub.conf**
```
  default 0[1]   ## boot from label 1/2
  timeout 10
  title Boot from disk 1
    rootnoverify (hd0,0)
    chainloader +1
  title Boot from disk 2
    rootnoverify (hd1,0)
    chainloader +1
```

# 6 STORAGE

## 6.1 Open issues in storage management

Digitizing the information stored in every book in the Library of Congress requires about 10 terabytes of disk space. Storing all the movies ever made on DVDs needs 5 petabytes of data. Five petabytes are also the amount of data the Large Hadron Collider of CERN expects to collect annually in coming years. Half to one petabyte is the space required to store all the virtual machine disks of a medium (1000 CPU) computing center.
High Performance Computing consists of three major subsystems: processing, networking and storage. While computing and network are often enough for the current requirements, storage is more and more a strong limitation to the overall system performance. Improving performance, capability and throughput are the biggest engineering challenges for the future of storage [45].

One of the side effects of putting a lot of data in a single computing center is the increase in aggregate bandwidth for the physical storage system, and the growing of I/O and RAM requirements for the disk servers providing access to the shared storage. Increasing by a factor N the computational power and the storage capability in a data center entails in many instances the increase of the aggregate bandwidth from/to the storage by a factor $N^2$.
Optimizing the storage layout can significantly help in the area of High Performance Computing and High Availability in virtual environment [46].

## 6.2 Storage media types

Storage refers to computer devices and recording media able to retain digital data, used for computing, for some interval of time. The computer storage provides the information retention. For this reason a safe system in absolutely needed, in order to prevent any loss of data.
Storage today commonly refers to mass storage: optical disks as CD or DVD, magnetic storage as hard disks, and solid state memory (more fast and expensive than other devices). The main feature of a storage system is the permanent nature of the saved data.
The magnetic hard disk uses different patterns of magnetization on a magnetically coated surface to save information. Magnetic storage is a kind of non volatile, permanent storage. The information is accessed using one or more read/write heads moving on the disk surfaces.

## 6.3 Magnetic storage

My dissertation concerns only the magnetic storage solutions: hard disk systems.
There are at least three magnetic permanent storage options available: Direct Attached Storage (DAS), Network Attached Storage (NAS), Storage Area Network (SAN).

## 6.4 Storage solutions: from NAS to SAN

Although the need for storage is evident, it is not always clear which solution is suitable for the organization at hand.

Choosing the right storage solution is not often a simple choice, depending on required data capacity, performance in aggregate bandwidth, and last but not least hardware and human costs. It is important to focus on the specific needs and long term business goals, taking into consideration several criteria:

- Capacity: the amount of data to be stored;
- Performance: I/O throughput requirements;
- Scalability: long term data growth;
- Availability and reliability: how critical are the applications?
- Data protection: backup and recovery requirements;
- Staff and resources available;
- Budget.

Fig. 6.1 [wpcontent.answers.com] shows the DAS, NAS, SAN Storage solutions.

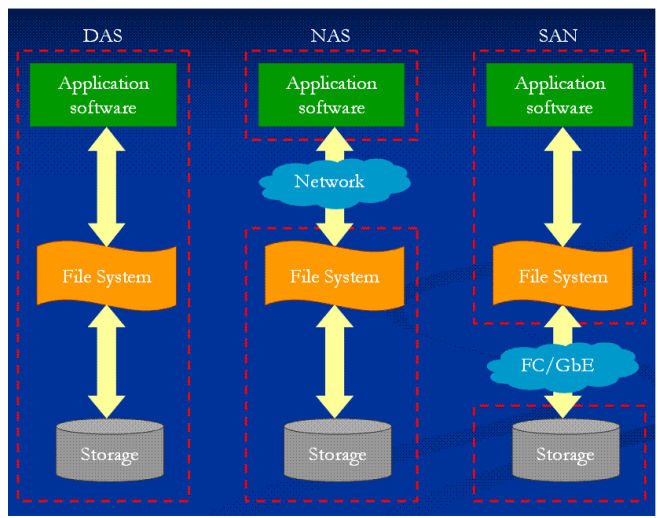

*Fig. 6.1: Storage solutions*

## 6.5 Direct Attached Storage - DAS

Direct Attached Storage -DAS is the simplest data storage connection system. Typically a Direct Attached Storage system is a disk drive or disk set directly attached to the server, and included in the same enclosure as the server. Direct-attached storage is the most basic level of storage, in which storage devices are part of the server providing the storage service.

Direct Attached Storage is usually connected to its server by a bus. There are many technologies used to connect the disk to the computer: the most common are SCSI, IDE/ATA, SATA and Fibre Channel.

This storage solution is typically made of one or more JBOD (Just a Bunch of Disks). Managing how the data on the disks are organized is under control of the server. This entails that in a Direct Attached Storage model a server has the dual functions of file sharing and application serving, potentially causing data serving slowdowns.
The Direct Attached Storage is the cheapest storage solution , and it can be a good solution for a relative small set of data, locally used, without a large I/O data requirement.
While a single Direct Attached Storage offers ease of management and administration, the management complexity increases quickly with the addition of new Direct Attached Storage servers, since the storage data space for each server must be administered and managed separately.

Fig. 6.2 [unixfoo.blogspot.com] shows a Direct Attached Storage - DAS storage architecture.

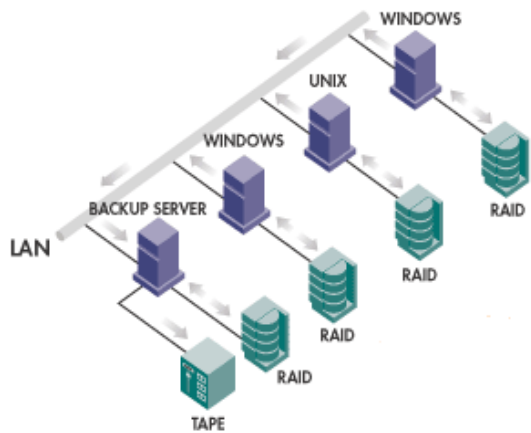

*Fig. 6.2: Direct Attached Storage - DAS*

## 6.6 Network Attached Storage - NAS

Network Attached Storage -NAS is one of the easiest and most cost effective way of adding shared storage capability to data center. A Network Attached Storage is a File Level Data Sharing Across the network.
By directly connecting a Network Attached Storage to an IP LAN, the storage is available to all the servers connected to the LAN, independently of the operating system types. On the network, a NAS system appears as a native file server to each of its different clients: Windows, Linux or other operating systems. This way the files are saved on or retrieved from the NAS system in their native file formats.
Network Attached Storage is a storage appliance to store and manage data access and retrieval through a set of attached dedicated servers, usually sharing the data via Network File System (NFS).
Network Attached Storage is a device comprised of both hard disks and management software, dedicated to serving files over a network, but applications independent. This way NAS provides a more flexible model in data access.

One of the main advantages of a Network Attached Storage system is that it is can be attached anywhere to the network, minimizing the costs and the impact on a computing center. NAS can be seen as a plug and play solution, cheep, easy to install and manage.
A single NAS system can provide many terabytes of storage in high density form factor, making efficient use of the computing and data center space.
The distribution over a data center of more dishomogeneous network storage devices can often limit the performance of the whole system, and thus the system scalability.

Fig. 6.3 [unixfoo.blogspot.com] shows a Network Attached Storage - NAS storage architecture.

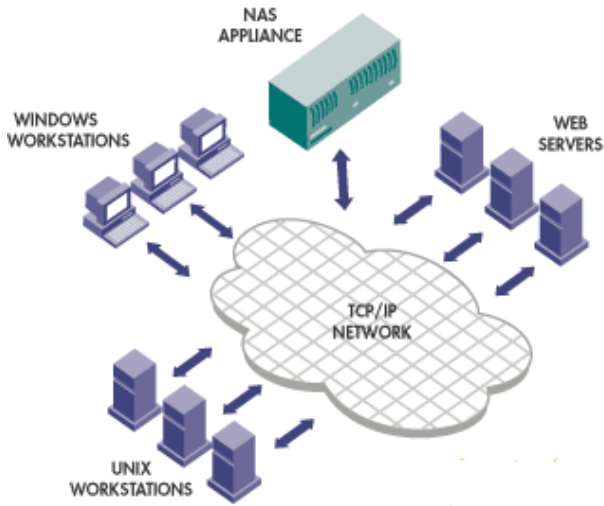

Fig. 6.3: Network Attached Storage - NAS

## 6.7 Storage Area Network - SAN

A Storage Area Network - SAN is a data storage architecture designed to meet the requirement of networking data storage behind the server.
The Storage Area Network solution is a dedicated, high performance storage network able to transfer data between servers and storage devices, separate from the local area network.
The Storage Network Industry Association SNIA definition of a Storage Area Network is: A network whose primary purpose is the transfer of data between computer systems and storage elements and among storage elements. A Storage Area Network consists of a communication infrastructure, which provides physical connections, and a management layer, which organises the connections, storage elements, and computer systems so that data transfer is secure and robust.

Often a Storage Area Network is used in connection with Fibre Channel (FC) technology, a solution able to increase the data transfer rate between disks and server. In a Storage Area Network infrastructure, more storage devices such as NAS, DAS, arrays or tape libraries may live together, connected to servers using Fibre Channel.

Fibre Channel has become the infrastructure of choice for Storage Area Network, for its ability to deliver large volumes of information in a rapid and efficiently way, and with predictable latency time. Fibre Channel is a highly reliable, gigabit interconnect technology that enables two way simultaneous communication among servers, data storage systems and other peripherals. Without the distance and bandwidth limitations of SCSI, Fibre Channel is ideal for moving a large volume of data across long distances in a quick and reliable manner.

With their high degree of sophistication, management complexity and cost, a Storage Area Network is traditionally implemented for mission critical applications and used in enterprise space or in high level research fields.

In contrast to Direct Attached Storage or Network Attached Storage, optimized for data sharing at the file level, the strength of a Storage Area Network architecture lies in its ability to move large blocks of data. This is especially important for bandwidth intensive applications.

The distributed architecture of a Storage Area Network is capable of offering higher levels of performance and availability than any other storage solution available today. By dynamically balancing data load and data transfer across the network via a multi path Fibre Channel infrastructure, Storage Area Network provides fast data transfer, reducing I/O latency and server workload. The main benefit is that a large number of users or applications can simultaneously access data without creating a bottleneck on the local area network and servers.

Storage Area Network is the best way to ensure predictable performance and 24x7 data availability and reliability, offering at the same time an excellent scalability.

Fig. 6.4 [unixfoo.blogspot.com] shows a Storage Area Network - SAN storage architecture.

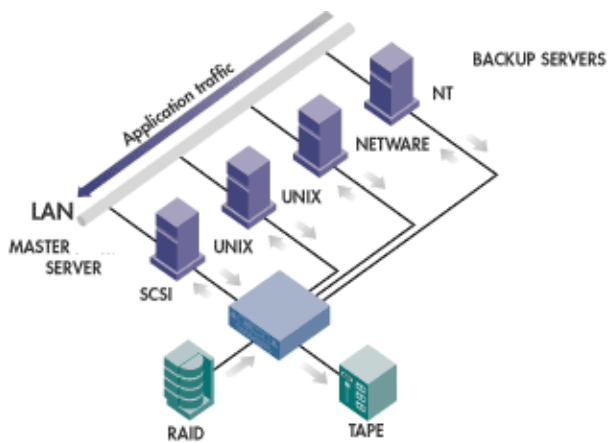

*Fig. 6.4: Storage Area Network - SAN*

## 6.8 iSCSI Storage Area Network - IP SAN

The iSCSI protocol is the encapsulation of the SCSI protocol within TCP/IP packets, mapping of SCSI (Small Computer Systems Interface) block-oriented storage data over TCP/IP networks. Developed by the Internet Engineering Task Force, iSCSI is a technology designed to carry SCSI commands and data across an IP network.

The iSCSI technology enables a server (initiator) to send SCSI commands and data to a storage device (target) over a TCP/IP based network in IP packets.

More recently iSCSI has emerged as an alternative which affects the more ubiquitous IP network. Using IP instead of Fibre Channel entails a large saving in cost and infrastructure project.

With the recent development of the iSCSI protocol, a Storage Area Network solution based on IP network is now possible. iSCSI joins the two most widely used protocols from the storage (SCSI) and the networking (IP) worlds.

On the storage side, iSCSI protocol uses the SCSI command set; on the networking side, iSCSI uses TCP/IP and Ethernet protocol, enabling application over a metropolitan and wide area networking as well.

Unlike some Storage Area Network protocols, iSCSI requires no dedicated cabling; it can be run over existing switching and IP infrastructure. As a result, iSCSI is a low cost alternative to Fibre Channel storage solutions, which requires a dedicated infrastructure.

Because SCSI is CPU intensive during high I/O load operations, iSCSI Host Bus Adapters (HBA) is a technology converting SCSI protocol into IP protocol, acting as a Fibre Channel HBA, except that Ethernet is used instead of Fibre Channel. The idea is that the SCSI requests are offloaded from the CPU onto the iSCSI HBA.

## 6.9 File systems

In computing technology, a file system is a method for storing and organizing files and the contained data to make it easy to find and access. A file system is used over a data storage device such as a hard disk in order to maintain the physical location of the files. A file system may also provide access to data on a file server by acting as clients for a network protocol (e.g.: NFS, SMB), or it may be virtual and exist only as an access method for virtual data (e.g.: procfs). More formally, a file system is a special purpose database for the storage, organization, manipulation, and retrieval of data on a storage device.

My dissertation analyzes only two of the network file systems, needed to share storage over a network storage infrastructure: NFS and GPFS

### 6.9.1 NFS - Network File System

Network File System (NFS) [47] [48] allows a system to share directories and files with others over a network. By using NFS, users and programs can access files on remote systems almost as if they were local files.

Network File System allows hosts to mount partitions on a remote system and use them as local file systems. This allows the system administrator to store resources in a central location on the network, providing authorized users a continuous access.

One of the most notable benefits that NFS can provide is that local machines may use less disk space if commonly used data are stored on a single machine and still remain accessible to other machines over the network. There is no need for users to have separate home directories on every network machine: users home directories can be set up on the NFS server and made available throughout the network.

Storage devices such as floppy disks, CDROM drives, pen drives, and Zip drives can be used by other machines on the network. This may reduce the number of removable media drives throughout the network.
NFS consists of two parts: a server and one or more clients. The client remotely accesses the data stored on the server machine. To function properly, a few processes have to be configured and run.

Fig. 6.6 [uic.rsu.ru] shows a Network File System - NFS architecture.

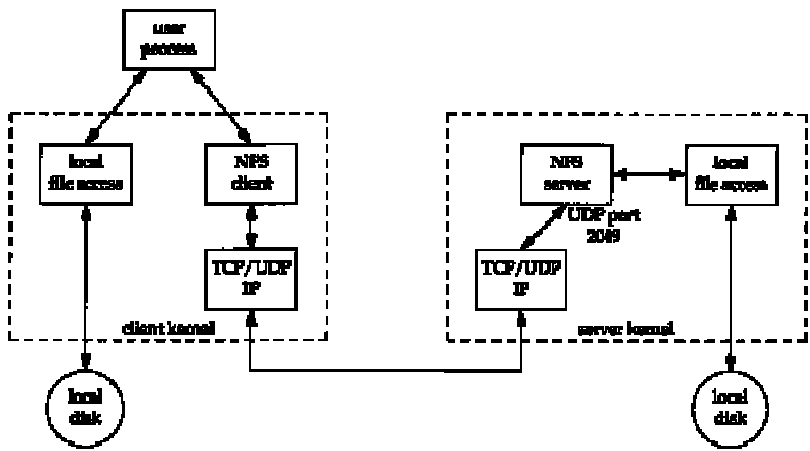

*Fig. 6.6: Network File System - NFS*

### 6.9.2 GPFS - General Parallel File System

The IBM General Parallel File System (GPFS) [49] [50] is a high performance scalable file management solution that provides fast, reliable access to a common set of file data from a single computer to hundreds or thousands of systems.
GPFS provides concurrent high-speed file access to applications executing on multiple nodes of clusters, and high I/O throughput by allowing data to be accessed over multiple computers at once.
Most existing file systems are designed for a single server environment, and adding more file servers does not improve performance. GPFS provides higher I/O performance by striping blocks of data from individual files over multiple disks, and reading and writing these blocks in parallel. Other features provided by GPFS include high availability, support for heterogeneous clusters, disaster recovery, security.

GPFS software has to be installed over all the file servers and clients, creating a cluster sharing data over the network.

Fig. 6.5 [ibm.com] shows a General Parallel File System - GPFS architecture.

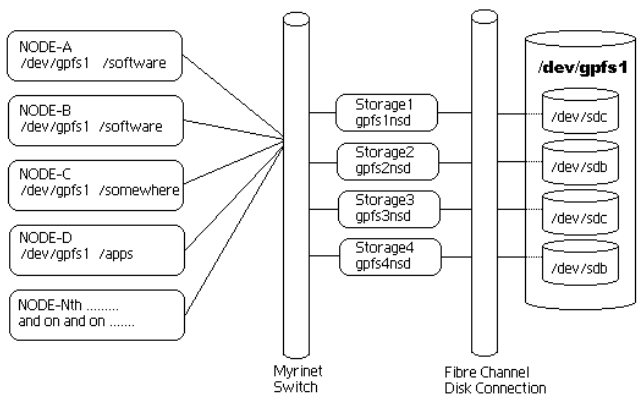

Fig. 6.5: General Parallel File System - GPFS

## 6.10 Grid Storage challenge

A redundant, secure, standardized Grid storage capability out-of-the-box is one of the most desirable technologies for both enterprise and scientific environments. If a centralized storage system can surely reduce the time spent in management and backup recovery, moving to a virtual infrastructure with disk space shared all around the world implies a new re-routing model and a new paradigm in data access.

## 6.11 Throughput Performance vs Data Safety

While RAID technology for servers has been created to protect data against hardware failure, some application is primarily focused on the possible performance gain.
Using faster and homogeneous storage technology (Fibre Channel, SCSI, RAID controllers) may enhance the throughput of a storage data system. Using smaller disks can increase parallelism, and thus the data transfer rate towards the disk JBOD. A solid state cache layer can improve local performance.

A multi path infrastructure is needed to guarantee more than one single path to reach the storage system. At the same time a multi path architecture may be used to increase parallel streaming towards the disk devices. The choice depends on the primary purpose to be achieved: data safety or throughput performance.
In a Grid architecture involving more than 1000 sites spread everywhere in the world, data movement and data access mode are at the present time two of the biggest problems to solve and a challenge for the future.

## 6.12 RAID technology

Redundant Array of Independent Disks, also known as RAID, is a system based on multiple hard disk drives for sharing or replicating data.
The main idea is to divide and replicate data among multiple hard disk drives. All the various designs of RAID involve two key design goals: increase data reliability and increase I/O throughput performance. When multiple physical disks are set up to use RAID technology, they are said to be in a RAID array. This array distributes data across multiple disks, but the array is seen by the system as one single disk. RAID can be set up to serve several different purposes, such as performance increasing, or data redundancy.

Data redundancy is achieved by some extra data written across the array. Data are organized so that the failure of one (sometimes more) disks in the array will not result in loss of data. A failed disk may be replaced by a new one, and the data are reconstructed from the remaining data and the extra data. A redundant array allows less data to be stored than the sum of all disks capacity.
For instance, a 2 disk RAID 1 array loses half of the total capacity that would have otherwise been available using both disks independently, and a RAID 5 array with several disks loses the capacity of one disk over all the available disks. Other RAID level arrays are arranged so that they are faster to write to and read than a single disk.

There are various combinations of these approaches giving different trade-offs of protection against data loss, capacity, and speed. RAID levels 0, 1, 5 and 6 are the most commonly used, and cover most requirements.

### 6.12.1 RAID 0

RAID 0 - STRIPE: distributes data across several disks in a way that gives improved speed and full capacity, but no data redundancy is available; all data on all disks will be lost if any disk fails. It provides improved performance and additional storage but no fault tolerance. Any disk failure destroys the array, and this becomes more likely with more disks in the array. A single disk failure destroys the entire array because when data is written to the array, the data is broken into fragments.

The number of fragments is dictated by the number of disks in the array. The fragments are written on the same sector of their respective disks simultaneously. This allows smaller sections of the entire chunk of data to be read off the drive in parallel, giving this type of arrangement a huge bandwidth. RAID 0 does not implement error checking so any error is unrecoverable. More disks in the array means higher bandwidth, but greater risk of data loss.

Fig. 6.7 [Wikipedia] shows a two disk RAID 0 system.

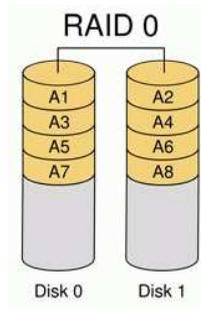

*Fig. 6.7: RAID 0 disk array*

### 6.12.2 RAID 1

RAID 1 - MIRROR: a real-time backup solution. Two disks store exactly the same data, at the same time, and at all times. Data is not lost as long as one disk survives. Total capacity of the array is simply the capacity of one disk.

Fig. 6.8 [Wikipedia] shows a two disk RAID 1 system.

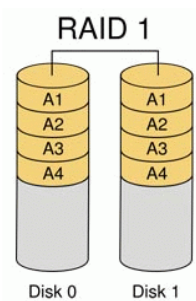

*Fig. 6.8 RAID 1 disk array*

### 6.12.3 RAID 5

RAID 5 - STRIPE WITH SINGLE PARITY: distributes a parity check over all the array disks; protects data against loss of any disk; requires all drives but one to be present to operate; a single drive failure requires replacement, but the array is not destroyed. Upon a drive failure, any subsequent reading can be calculated from the distributed parity such that the drive failure is masked from the end user.

The array will have data loss in the event of a second drive failure and is vulnerable until the data that was on the failed drive is rebuilt onto a replacement drive. The storage capacity of the array is reduced by one disk.

Fig. 6.9 [Wikipedia] shows a RAID 5 disk array.

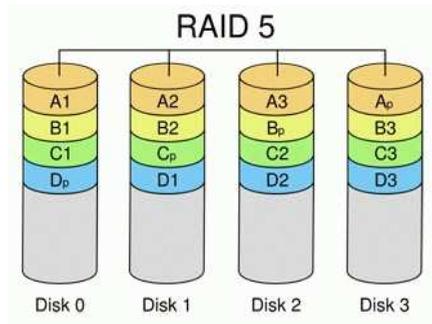

*Fig. 6.9: RAID 5 disk array*

### 6.12.4 RAID 6
RAID 6 - STRIPE WITH DOUBLE PARITY: distributes two parity checks over all the array disks; provides fault tolerance from two drive failures; array continues to operate with up to two failed drives. This makes larger RAID groups more practical, especially for high availability systems. This becomes increasingly important because large-capacity drives lengthen the time needed to recover from the failure of a single drive.

Single parity RAID levels are vulnerable to data loss until the failed drive is rebuilt: the larger the drive, the longer the rebuild will take. Dual parity gives time to rebuild the array without the data being at risk if one drive, but no more, fails before the rebuild is complete.

Fig. 6.10 [Wikipedia] shows a RAID 6 disk array.

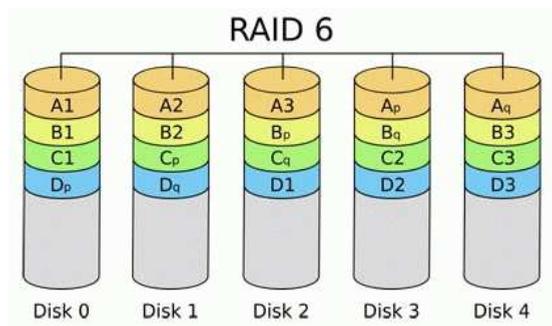

*Fig. 6.10: RAID 8 disk array*

### 6.12.5 Other RAID
- RAID 0+1: striped sets in a mirrored set (minimum four disks; even number of disks) provides fault tolerance and improved performance but increases complexity. The key difference from RAID 1+0 is that RAID 0+1 creates a second striped set to mirror a primary striped set. The array continues to operate with one or more drives failed in the same mirror set, but if drives fail on both sides of the mirror the data on the RAID system is lost.

- RAID 1+0: mirrored sets in a striped set (minimum four disks; even number of disks) provides fault tolerance and improved performance but increases complexity. The key difference from RAID 0+1 is that RAID 1+0 creates a striped set from a series of mirrored drives. In a failed disk situation, RAID 1+0 performs better because all the remaining disks continue to be used. The array can sustain multiple drive losses so long as no mirror loses all its drives.

- RAID 5+0: stripe across distributed parity RAID systems.

- RAID 5+1: mirror striped set with distributed parity.

## 6.13 RAID notes

RAID can involve significant computation power when reading and writing data. With traditional RAID hardware, a separate controller does this computation. In other cases the operating system requires the host computer processor to do the computing, and this reduces the computer performance on applications highly CPU intensive.

RAID systems with redundancy continue working without interruption when one or sometimes more disks of the array fail. When the bad disk is replaced by a new one, the array is rebuilt while the system continues to operate normally.

## 6.14 Mirror over network - Distributed Replicated Block Device

The Distributed Replicated Block Device (DRBD) [51] is a software-based, replicated storage solution mirroring the content of block devices (hard disks, partitions, logical volumes, etc.) between servers over the network.

DRBD mirrors data in real time. Replication occurs continuously, while applications modify the data on the device.

Transparently to the end user, the applications store their data on the mirrored device, duplicating stored data on several computers.

Data mirror can be performed synchronously or asynchronously. With synchronous mirroring, a writing application is notified of write completion only after the write has been carried out on both computer systems. In asynchronous mirroring the writing application is notified of write completion when the write is completed locally, but before the write has propagated to the peer system.

The DRBD core functionality is implemented by way of a Linux kernel module. Specifically, DRBD constitutes a driver for a virtual block device, so DRBD is situated right near the bottom of a system's I/O stack. Because of this, DRBD is extremely flexible and versatile, which makes it a replication solution suitable for adding high availability to any application.

Fig. 6.11 [drbd.org] shows a Distributed Replicated Block Device - DRBD system.

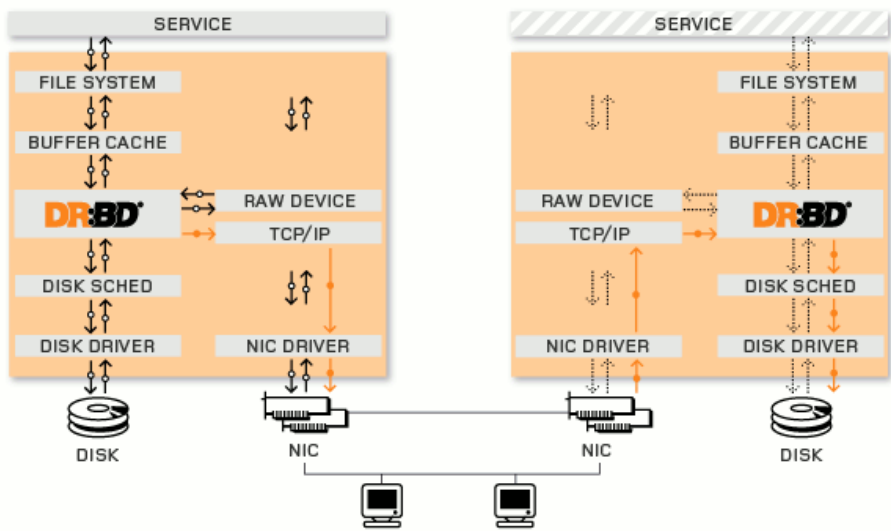

Fig. 6.11: Distributed Replicated Block Device - DRBD

## 6.15 Case studies and solutions

A set of experimental data taken during the last five years at the Italian National Institute of Nuclear Physics INFN and Scuola Normale Pisa, several architectural storage solutions have been tried, and a lot of storage performance tests have been carried out.
While Direct Attached Storage DAS can be useful for a small set of data shared from a single server, Network Attached NAS and Storage and Storage Area Network SAN are the best choice for a large set of data to be shared over a large number of hosts. The use of NAS rather than SAN depends essentially on the available budget. At the moment a single NAS system can reach an I/O throughput of 4 Gb/s, while a complex - and too expensive - SAN can reach an I/O throughput of 100 Gb/s.

A multi path infrastructure to reach the disk array has been deployed with a set of redundant switching systems. This way a failure of a single switch does not affect the correct operation.
With respect to data redundancy and reliability, the chosen solution has been almost always the RAID 6 technology, due to the high capability of the single disks today. In case of a very limited budget, a Distributed Replicated Block Device DRBD has been developed to reach the same data reliability with no additional costs.
A great attention must be payed for the network configuration, in order to guarantee the required I/O data throughput and to avoid bottlenecks between disk devices and servers.

### 6.15.1 GPFS over SAN + RAID 6

At Italian National Institute of Nuclear Physics INFN Pisa a computing center has been created in the last three years for the study of Elementary Particle Physics. The computing center comprises today more than 2.000 CPU, 500 TB disk and a full 1 GB switching infrastructure.

A Storage Area Network with double redundant Fibre Channel head, four GPFS server with 10 Gb network card on board and four 2 Gb Fibre Channel each one, are used to reach 16 Gb/s I/O sustained throughput. This rate has been held for a 24 hours stress test without any failure in the data chain. RAID 6 has been the choice to guarantee a good data reliability without performance reduction.
This storage architecture can be seen as a high level cost solution, not recommended in case of low budget project.

### 6.15.2 NFS over DAS + DRBD

At Scuola Normale I developed my research work on High Availability using virtual environment. Given that the main goal on my research has been to reach a satisfactory availability level with the minimum expense, a low cost storage solution has been studied.

A mirror over network, exploiting a Distributed Replicated Block Device DRBD, has been developed to reach the same data reliability of a Storage Area Network with multi path with no additional costs. The main differences between this solution and a more expensive Storage Area Network solution are the data capacity (1 TB of data vs 500 TB), the I/O data aggregate throughput (30 MB/s vs 2 GB/s), the system scalability.
Such a solution has been considered suitable for the resources demand of the project.

## *6.16 Conclusions*

With such a variety of storage technologies available, the best way to determine which one - DAS, NAS, or SAN - is the best choice for the required application field is to investigate pro and contra of every storage solution.
The first step is to consider the nature, the amount, and the transfer rate of the data and applications, how critical they are, and what are the minimum required levels of performance and availability. After that, another important characteristic to investigate is the data localization or distribution across the network.

An optimal storage configuration is not an easy choice. A lot of factors have to be taken into consideration: how many TB per server? which RAID configuration? A fine tuning of parameters is needed in disk arrays, controllers and servers (cache, block sizes, buffer sizes, kernel parameters). With respect to the storage pool architecture: is one disk pool large enough, or is it necessary to split?
The optimal configuration depends strongly on the application: for a data intensive application, an important role is played by the size and the rate of data to be transferred. Different file systems have been developed to improve data transfer in case of small or large data sets.

Disaster recovery is today the top of data storage challenges. A cheap solution will be shown in the next chapters.

# 7 PROPOSAL AND SOLUTIONS

## *7.1 Background and motivations*

One of the most critical issues for a computing center is the ability to provide a high availability service for all the main applications. Today all services are intended to be by definition 24x7 (24 hours at day, 7 days per week), even those with a short life span.
The goal of all availability systems is to maximize the uptime of the various online systems for which they're responsible, to make them completely fault tolerant.
There are several approaches able to maximize availability of a single service running on a single host. The real problem is originated by the needs of high availability of a whole computing center, running many services over heterogeneous operating systems.

A lot of business solutions exist at the moment to satisfy the needs of service availability in production environment. The problem is often related to the low portability and flexibility of the solution. On top of that, a lot of proprietary solutions are not free, with a cost often too high for a research computing center or for a small company.
By analyzing the causes of planned and unplanned downtime of a large computing center (INFN Pisa: 2000 CPU, 500 TB disk) for a period of three years, a new approach to high availability has been supplied, based on the features of virtual machine running on virtual environment [52].

## *7.2 High availability using virtualization*

Since a not critical availability level is often required by almost all services running on web, Grid, or other technological environments, an interesting solution is offered by a new availability idea: a system able to restore an application in a few minutes. Usually a 2 to 3 minutes blackout is not critical for most part of the production applications. By paying for the non immediate recover of a crashed application, a new approach to high availability is offered by intensively exploiting all the features of a virtual environment.

The idea is to satisfy the research and the enterprise needs of high availability with a zero cost new solution, based on the idea that a relaxed system may ensure the application redundancy required in the greater part of the cases. "Relaxed" means a system able to restore any previously running application in less than ten minutes from the crash time. While usually high availability services as cluster or heartbeat are able to recover a failure in a few seconds, a relaxed system is a solution able to restore a service in a few minutes.

The originality of this new approach to high availability is that a computing center system administrator has not to worry about the system redundancy till the disaster occurs. Only at that moment the system is able to restore the crashed application in a location result of a choice algorithm. 3RC is the name of the project, acronym for 3 R Cycle [53].

The aim of my work is to provide a high availability system - architecture, software, methodology - for all non critical services, without charging a computer center with the cost of the classical heartbeat host per host redundancy.

High availability virtual infrastructure ensures that a service has constant availability of network, processors, disks, memory. This way a failure of one of these components is transparent to the application, with a maximum time delay of two to ten minutes.

Fig. 7.1 [thanks to Claudio Atzeni] shows the 3RC logo.

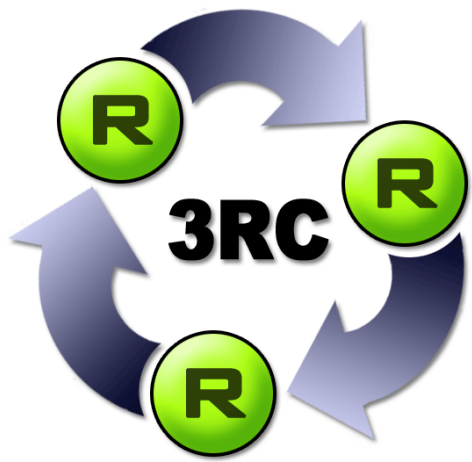

*Fig. 7.1: 3RC logo*

Using virtualization, it is possible to achieve a redundancy system for all the services running on a data center. This new approach to high availability needs only one remote controller and one backup server - also shared among the hosts in the computing center - running a virtual environment, to recover services usually distributed on several physical hosts.
Exploiting virtualization and ability to install a host from scratch in a few minutes, it is possible to do a sort of host on-demand, where the start-up of a backup virtual machine is done only when the disaster occurs.

## 7.3 Redundancy in virtual environments - 3RC

The 3RC system offers several possibility of redundancy strategies, depending on the desired high availability level. There is not a unique solution for all possible requirements; each use case needs a different processing solution.
The system is able to recover all the managed services running in virtual environment in case of crash or failure. The unavailability of the original physical host providing the virtual machine - e.g. due to a physical machine fault - is managed in an absolutely transparent manner.

Dividing host crashes into operating system and applications faults, my work covers the only operating system crash case.

### 7.3.1 Operating system level

The host status (up or down) is determined by the host operating system response by checking the operational state of a set of specific applications, such as ping, health monitor state. The 3RC system is able to recover only a crashed operating system status.

### 7.3.2 Application level

No check is performed at application level. The recovery of a crashed application in a working operating system has to be managed by the application itself, or rather by a controller script running on the same host. It is a programmer's task to ensure that an application is able to check its correct operational status. Usually this is achieved by an extra controller scheduled to periodically run in background; this script checks the application response via simple controls, and must be able to perform operations such as the application start or stop.

## 7.4 Architecture

The solution has been improved in a virtual environment by using a remote controller and an application based on a finite state machine. Each state plays the role of an action which can be performed on the single virtual machine hosting the crashed application or on the whole virtual layer running over the physical hosts.
Moreover a zero cost storage redundant solution has been deployed in order to guarantee a certain level of data reliability and availability in case of damage of one system device, and to reduce the possible bottlenecks.
The whole PXE infrastructure described in a previous chapter has been used to allow a virtual machine to be installed in a few minutes from scratch.

## 7.5 Controller

The remote controller is the main core of the 3RC high availability system. It is responsible of checking and monitoring all the hosts involved in the high availability service.
The only requirement for the controller is the ability to access as administrator all the involved hosts. Usually this can be performed by a public key exchange between the controller and each node of the cluster, in order to guarantee the access without the password request by the accessed host in case an operation is required.

## 7.6 Virtualization layer

An accurate preliminary study of all available virtualization architecture has been performed. The evaluated solutions has been essentially XEN and VMware.
While XEN is a complete free and open source software, VMware is a business closed source solution, with a free release available.
My end choice for the virtualization layer has fallen to the VMware free server solution. The reasons for that choice have been the VMware business reliability and the guarantee of a continuous software package update.

## 7.7 Operating systems

While the main controller of the high availability system needs a Linux based host, the only requirement for the servers hosting the virtual machines is to run the VMware server. The virtual machines can run whatever kind of operating system, from Unix to Windows.

## 7.8 Infrastructure

A complex infrastructure, based solely on servers ever available in a typical computing center, has been developed to install operating system and middleware in a few minutes, by exploiting Preboot Execution Environment PXE technology, as well as DNS and DHCP services. Using this infrastructure more than 5000 servers have been installed from scratch by the knowledge of their only MAC address.
A redundant controller system periodically checks all the virtual machines running in a high availability status.

## 7.9 Storage

Several storage solutions, from Network Attached Storage NAS to Storage Area Network SAN have been tested to store and centralize all the virtual disks, to grant data safety and access from everywhere, enhancing the aggregate bandwidth and reducing at the same time the downtime period. Depending on the required availability level, the financial resources, the storage redundancy method, a lot of possible solutions are available.
The Storage Area Network solution is certainly the best solution in terms of performance, reliability, availability, but the cost of such a solution is generally prohibitive for a research institute or university. While today the cost per TB is about 100 € for low cost disk, it can increase till 10.000 € per TB in case of a high performance Storage Area Network. For the Network Attached Storage, the best solution in my opinion for a medium computing center, the cost per TB is between 500 and 1.000 €.
A zero cost solution - the adopted one in my research - is played from the DRBD mirror over network.

### 7.9.1 Mirror over network

The Distributed Replicated Block Device (DRBD) is a software-based, replicated storage solution able to mirror a whole disk, a partition or a logical volume between servers across the network. DRBD mirrors data in real time, in a transparent way to the end user, duplicating stored data on several computers.
This is performed by adding a module to the Linux kernel.

## 7.10 Monitor service

A lot of monitoring systems have been tested to check the hosts status. Even if someone of them is able to perform action on the analyzed machines, the chosen solution has been the Ganglia Monitoring System [54].
Ganglia is a distributed monitoring system for high performance computing systems such as clusters and Grids. For this reason it reasonably meets the system requirements: check and report any host failure in an acceptable time.

Ganglia is a scalable distributed monitoring system for high performance computing systems such as clusters and Grid. It is based on a hierarchical design and uses technologies such as XML for data managing, and RRDtool for data storage and visualization. The system uses very light data structures and algorithms to achieve very low per node overheads and high concurrency.

The tests performed over five years have proved that the implementation is robust enough: more than 99.9% of the crashed cases are discovered and announced in an average time of 70 seconds. The default polling interval (the hosts check sampling interval) is 15 seconds.
An additional plugin has been developed to make Ganglia information available for a Linux shell script. It periodically analyzes the XML Ganglia log in order to manage the information in a more flexible and usable way.

Fig. 7.2 [SNS] shows a typical Ganglia monitor screenshot.

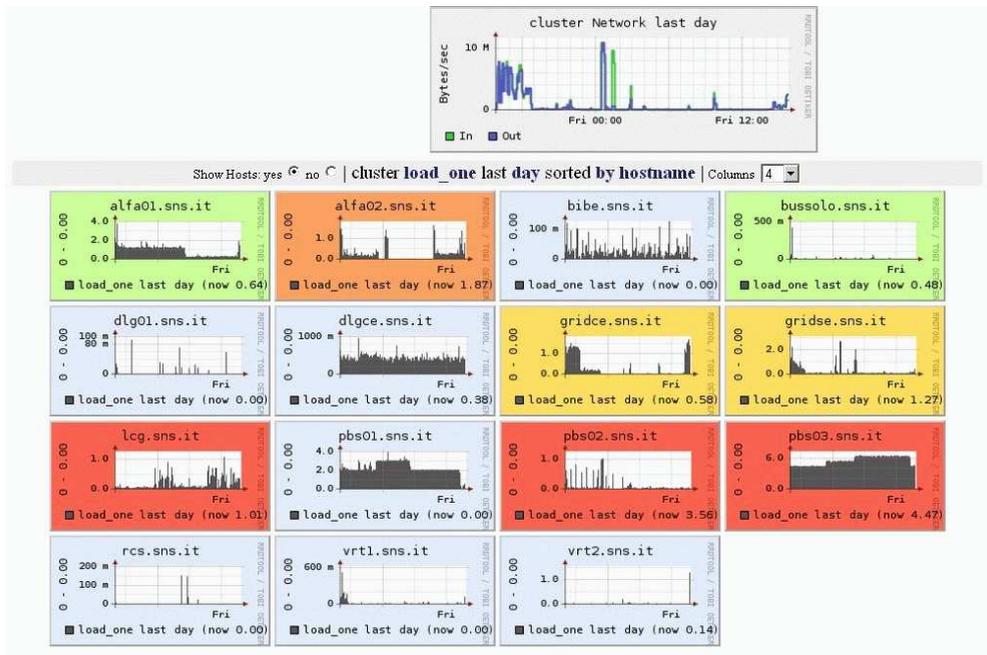

*Fig. 7.2: Ganglia monitor screenshot*

The developed plugin extracts in text format all the needed information related to physical and virtual host, to be passed to the remote controller 3RC. The extracted information are: host name, average load in the last 5 minutes, seconds from the last ping received.

```
HOSTS STATUS

hostname;load;last_ping;

alfa01.sns.it;2.94;25;
alfa02.sns.it;1.13;18;
alfa03.sns.it;0.01;24;
bibe.sns.it;0.01;13;
bussolo.sns.it;0.00;20;
gridce.sns.it;1.55;15;
gridse.sns.it;0.28;26;
lcg.sns.it;0.00;13;
pbs01.sns.it;2.00;13;
pbs02.sns.it;1.04;19;
pbs03.sns.it;5.74;21;
vrt1.sns.it;0.00;15;
vrt2.sns.it;0.00;28;
```

## 7.11 Time schedule

A control script (the remote controller) is scheduled to run every 60 seconds on a single host - the host can be both internal or external to the cluster. 70 seconds is the average time needed from Ganglia to be aware of a machine operating system fault. The consequent average total time from a machine crash to its awareness by the 3RC controller is 30 + 70 seconds, with a confidence interval of ± 30 seconds. It means 100 +/- 30 seconds needed to be aware of a machine crash.

## 7.12 Finite state machine approach

The intervention level on a crashed machine, in 3RC project, goes from a simple host reboot till a complete operating system and middleware re-installation, through a restart service.
A finite state machine with cyclic states has been developed to meet the required intervention levels. Sequentially switching the three provided states, in a sort of intervention escalation, it allows to operate in three different manners on the crashed system. The three levels provided are: reboot, restart, reinstall.

The remote controller scans once per minute the whole cluster. In case of a failure detected, it checks the previous state of the virtual machines involved in the failure, and acts consequently. Only if the reboot or restart procedures have already been tried, the reinstall procedure starts.
For a typical Linux environment, a 3 minutes time wait between the reboot or restart and the reinstall procedure has been setup.

### 7.12.1 Reboot

The first and simplest solution for a high availability service is to reboot the machine when an operating system crash occurs. It could be very difficult to achieve: the reason is that if an operating system crashes, with a large probability it is not contactable from outside. In case of machine isolation, the reboot cannot be induced from outside.

The reboot intervention level, in case of virtual machines running over a physical host, is almost unworkable. It can be easily removed from the action list to be performed on a failed host.

### 7.12.2 Restart

The term "restart" in my work means the restart of the virtual layer managing the crashed virtual machine. This way is it possible to manage a virtual machine switch on and off, as a real one. While a human intervention is needed to switch on a turned off physical host - with the exception of servers provided with remote hardware controller listening on the network - the switch on a turned off virtual machine can be done also from a remote controller.
By acting a restart of the virtualization layer managing the virtual machine, the system can easily operate a complete restart of the virtual machine, including the beginning bootstrap procedure.

The restart of a virtual machine can be performed on the same physical host previously hosting the virtual machine, or on another physical host, depending on a choice algorithm based on:
- physical hosts availability;
- average load of physical hosts in the last 5 minutes;
- number of virtual machines already hosted by the single physical host.

A threshold load level is given for each physical server, depending on its CPUs number, RAM availability, internal processor architecture.
This way, if a physical host serving one or more virtual machines crashes, the hosted virtual machines can be easily moved to another physical host as soon as the controller detects the failure (the control check is scheduled once per minute). If no one of the available physical hosts in the cluster are at the moment free, the virtual machine waits to be restarted till a physical host reaches an under-threshold load level.

### 7.12.3 Reinstall

As a last chance, the system can provide a complete reinstall procedure, exploiting the whole PXE infrastructure previously described. After the reboot and / or the restart procedure have been performed unsuccessfully, the remote controller changes the link between the crashed virtual machine MAC address and the installation configuration required in the tftpboot/pxelinux.cfg path. The information about which operating system and middleware are required for a specific host are provided by a configuration file.

After the restart of the virtual machine, at the bootstrap time, the new kernel is loaded and the new installation starts from scratch.
The reinstall procedure can be disabled on a set of virtual machines, depending on the specific requirement - e.g. in case of non automatic installation provided for that kind of host.
As for the restart procedure, the reinstallation can be performed on the same or on another physical host, depending on the physical host availability, average load and status.

## 7.13 Recovery time

More than 15.000 crash test have been performed over 4 different testbed during a two week stress test period. Both the host and the guest operating system used are Linux: Debian the host, Red Hat the guest one.
Two crash cases will be analyzed:
- non destructive crash;
- destructive crash.

### 7.13.1 Non destructive crash

For a simple non destructive crash the chosen procedures have been:
- simple switch off: the virtual machine is switched off by a halt, or shutdown command;
- high load increase: a recurrent procedure aimed at increasing the virtual machine load until the system crashes or does not respond.

The time for a virtual machine Red Hat operating system Reboot is about 80 seconds [10 seconds for the PXE network setup procedure + 70 seconds for the boot] $\pm$ 10 seconds
The boot time added to the time the controller needs to recognize a failure gives the total recovery time in case of non destructive crash: 180 seconds $\pm$ 30 seconds.
The Gaussian (normal) distribution fitting the observational data has a mean of 181 seconds and a standard deviation (sigma) of 10 seconds.

Fig. 7.3 [experimental data] shows the recovery time after 10.000 non destructive crashes.

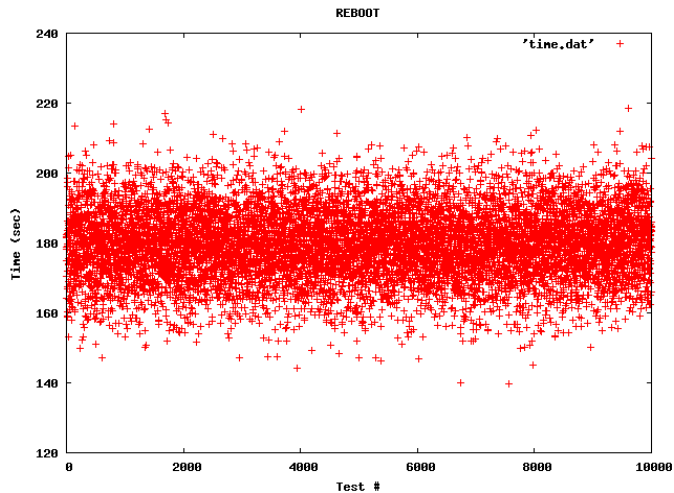

Fig. 7.3: Recovery time - 10.000 crash test

Fig. 7.4 [experimental data] shows the recovery time distribution after 10.000 non destructive crashes.

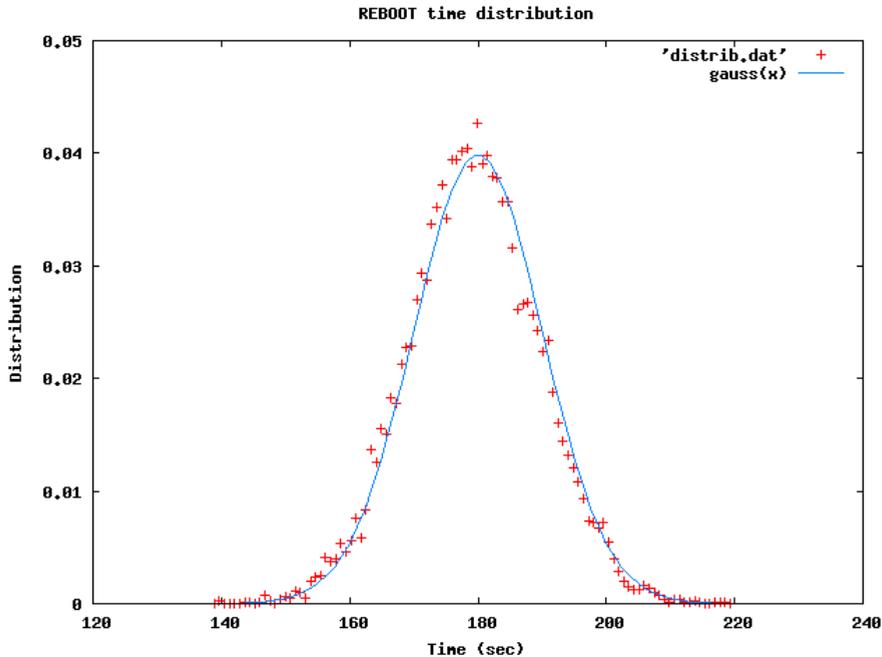

*Fig. 7.4: Recovery time distribution - 10.000 crash test*

### 7.13.2 Destructive crash

For a destructive crash the chosen procedure used has been a reboot after the /boot partition has been erased. This is, in my opinion, the only safe way to certainly destroy a Linux based system.

The time for a virtual machine Red Hat operating system Reinstall is about 442 [10 seconds for the PXE network setup procedure + 352 seconds for the installation + 80 seconds for the boot] ± 17 seconds
The install time added to the time the controller needs to recognize a failure gives the total recovery time in case of destructive crash: 542 seconds ± 45 seconds.
The Gaussian (normal) distribution fitting the observational data has a mean of 542 seconds and a standard deviation (sigma) of 17 seconds.

Fig. 7.5 [experimental data] shows the recovery time after 5.000 destructive crashes.

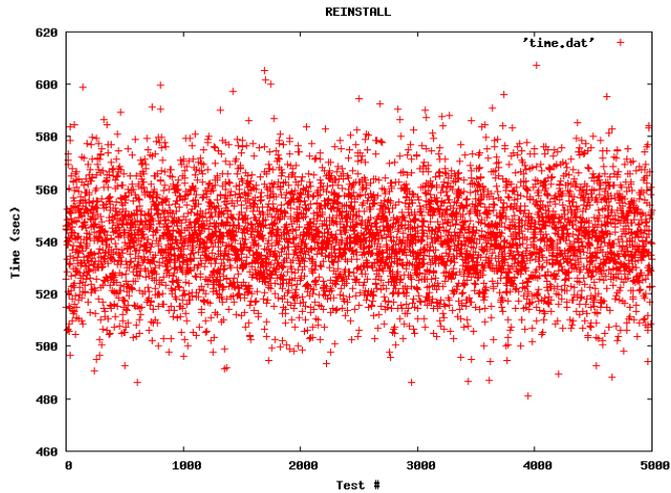

*Fig. 7.5: Reinstall time distribution - 5.000 destructive test*

Fig. 7.6 [experimental data] shows the recovery time distribution after 5.000 destructive crashes.

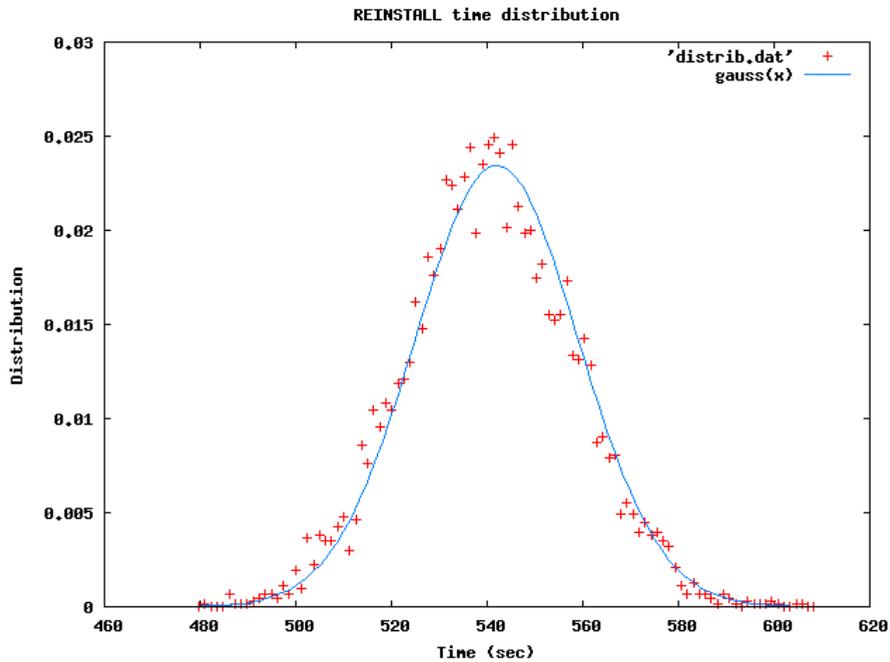

*Fig. 7.6: Reinstall time distribution - 5.000 destructive test*

## 7.14 Advantages

By using an automated system to reinstall from scratch a damaged host, the administrator no longer has to physically put hands on the software or hardware problems when a computer crashes. A network boot does the reinstall in a complete automatically way.

In a lot of cases, it is more efficient to reinstall operating system and software than to try to determine the problem with the existing installation.

By booting from the network, you get a guaranteed clean operating system, with no viruses or user modified files.

## 7.15 Physical to virtual migration

To virtualize the main servers of a data center, the first requirement to satisfy is the ability to virtualize a single host.

At Scuola Normale and INFN Pisa, in the last years I developed a new procedure for a semi-hot swap physical to virtual migration. The procedure provides:
- physical to virtual host migration. with a system downtime lower than 30 seconds in production environment;
- virtual to physical migration (vice versa) - never used at the moment.

The implemented procedure steps for a Physical to Virtual migration are essentially based on:
- virtual machine creation: via virtual layer cloning or setting up from scratch;
- synchronizing tool to hot synchronize original physical and new virtual host;
- new sync with both physics and virtual host down, in order to have a new system completely up to date.

The physical to virtual migration procedure has already been tested and used in LCG CERN Computing Grid environment.

# 8 OPERATION

## 8.1 Environment

The 3RC high availability service is aimed at taking care of services running over a computing cluster - a set of physical hosts serving more virtual machines in virtual environment.
Ganglia is the monitor used to check the machine (both physical and virtual) health status. A redundant remote controller - the core of my project - attends to the system overview, and manages the actions to be taken in case of failure of one or more of the system components.

## 8.2 Hysteresis based finite machine state

Hysteresis: tendency of a system to respond differently to the same stimulus depending on the initial state of the system. [definition by Claudia Guida, Molecular Biologist @ IEO Milan].

A system with hysteresis can be summarized as a system that may be in a finite number of states, independent of the inputs to the system. A system with hysteresis exhibits path dependence. It' s not possible to predict the output without looking at the history of the input - i.e. the state of the system for a given input. In order to predict the output, the algorithm must look at the path followed by the output before it reached its current value.

From this point of view, the choice algorithm of the 3RC core is able to perform an action - among the possible - based on the knowledge of the current and the previous states of the failed host. The algorithm, based on a finite state machine with hysteresis, is responsible of the action to be performed on each failed virtual host, in order to recover in the lowest time period the failed machines.

The steps performed from the remote controller are:

- **REBOOT**: simple reboot - in case of previous correct operational status of the virtual machine.

- **RESTART**: restart of the virtual layer managing the virtual machine (similar to a turn off and turn on of the virtual machine) - in case of an already reboot action performed in the last hour - the restart can be performed on the same physical host or on another one if the load of the original physical server is higher than a selected threshold level.

- **REINSTALL**: complete operating system and middleware reinstallation - this action is performed (only if expected in the configuration file) as the last choice, and only if all the previous action have already been performed.

It is possible to set up a variable number of cycles for each action to be performed before passing the control to the next action step.

Fig. 8.1 shows the 3RC finite state machine with hysteresis.

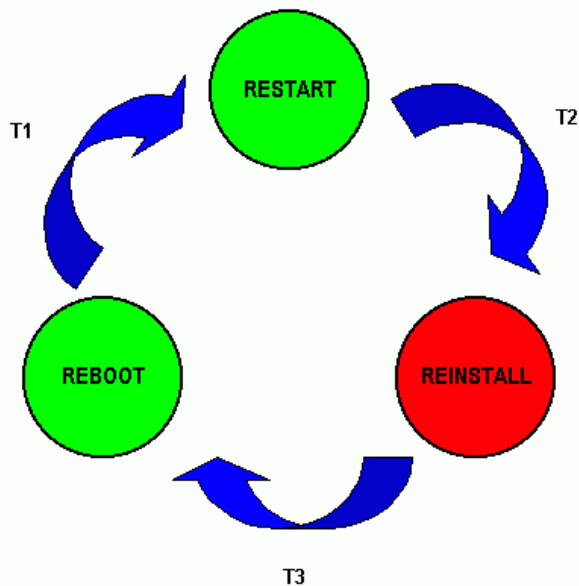

Fig. 8.1: 3RC finite state machine with hysteresis

## 8.3 Operational steps in case of host failure

What happens when a physical or virtual host crashes?
When a host fails, the controller is able to detect the failure in less than one minute after the Ganglia monitor event detection. A choice algorithm is responsible to decide on the best solution to be provided in order to recover the failed host(s).
As a consequence of a typical crash event - e.g. an overload of the virtual machine running a specific service - the monitor detects the failure 70 seconds after the failed host becomes unresponsive. The controller notices the failure with a delay of 0 to 60 seconds and starts the recovery procedure.

If the previous host state (the state of the virtual machine at the previous check, usually one minute before) was "up and running", the system try to do a reboot of the virtual machine involved in crash, hoping the machine is still reachable via ssh. After that, the system is in a sort of standby state, waiting till the end of T1 (3 minutes).
At the first check scheduled after T1, if the failed host is still unresponsive and no feedback comes back, the controller performs a restart of the virtual layer involving the failed virtual machine; this is equivalent to physically turn off and on the host, in order to restart operating system and all the loaded modules and applications.
If after a T2 (3 minutes) waiting time the host does not become responsive, compatibly with the system configuration permissions, the controller starts the complete reinstall procedure, performing a full installation from scratch: disk partition, disk formatting, operating system installation, post configuration process, middleware installation and configuration, services start up.

Each of the previous procedures can be performed a configurable number of time. In my opinion, the reboot attempt can be omitted; best practice could be to study the use case to determine if a single reboot can be performed or not - usually if an application crashes with a system overload, the whole system becomes unresponsive to any external signal, included a reboot command.

At each operation preformed by the system on a virtual machine, a log is written on a shared area.

After one hour of proper operation, all the logs are cleared, and the normal operation status is restored for the hosts involved in a previous failure.

### 8.3.1 Virtual machine failure

When a virtual machine fails - e.g. in case of system crash, software overload, virtual device failure - the remote controller, after the detection - via monitor - of the host failure, performs one of the three possible steps in order to recover the host. Based on the previous state of the virtual machine (the information is stored on a shared log file), the controller decides which action is the best one at the moment: reboot, restart, or reinstall.

### 8.3.2 Physical machine failure

In case of physical machine failure - e.g. broken hardware - the system detects, based on a choice algorithm, the most appropriate host in the cluster to recover the hosted virtual machines. The virtual machines involved in the physical machine crash are moved to a backup physical host in the cluster.

## *8.4 Remote controller*

The remote controller - the 3RC core - is the main actor of the project. It has to be guaranteed the access without any password requirement to all the hosts (physical and virtual) involved in the cluster running the high availability service. The controller can run on whatever physical host, both as user or administrator.

A single controller can be responsible for the health status of all the nodes of a cluster, or a double control may be deployed in order to prevent a single point of failure - even if a secondary point of failure: in case of controller failure the whole cluster continues its usual operation, and only a joint failure of host and remote controller can affect the correct operation.

## *8.5 Time schedule*

A remote controller performs a complete scan of all the hosts involved in the cluster at a scheduled interval period of 60 seconds.

After each action taken from the controller, a configurable inactivity period is expected to prevent a repetitive action, until a feedback is returned from the crashed host: 3 minutes after a reboot or restart of the virtual layer are waited before performing the next action step; 10 minutes is the maximum time period expected for a complete reinstallation of the host. In the meanwhile the controller is in standby, and no actions are performed over the cluster.

After one hour when non action are performed over a single virtual machine, the virtual machine status is reset to the original one - the correct operation one.

## *8.6 Redundancy in 3RC system*

### 8.6.1 Monitor

Ganglia monitor provides an intrinsic redundancy strategy: the monitor information is natively shared among several hosts belonging to the cluster. By setting the optional parameters in gmetad.conf (collector configuration file) and gmond.conf (client to be monitored configuration file) the Ganglia information are stored in more than one collector node.

**/etc/gmetad.conf**

```
# List of machines this gmetad will share XML with
  trusted_hosts 127.0.0.1 192.168.1.10 my.node.org
```

**/etc/gmond.conf**

```
# Enable any host connected to the gmond XML to receive data
  all_trusted on
```

### 8.6.2 Controller

A new approach in redundancy systems used to control the health state of a cluster has been developed, in order to avoid the service unavailability in case of the control failure.
The strategy lies in doubling the controller service by running the same control service on two different nodes. The idea is to set up the two controllers time schedules at alternating time steps, in order to guarantee an unique performed action at a certain time. Considering that a single controller action does not require more than 10 seconds, a time rotation of 1 minute is an appropriate alternating time.

This way the controller A acts at the even minutes, and the controller B acts at odd minutes. Both the controllers A and B write the performed actions on a common shared log area, independently from the action performer.
As a consequence of such an architecture, for the single controller is not interesting if the other controller is dead or alive; it can act independently. If one controller fails, the only effect on the system is the scheduled check frequency halving, passing the period by 1 to 2 minutes. It is acceptable for my production environment, and compliant with specification requirements.

### 8.6.3 Storage

The storage has to be redundant and reliable by default. To achieve a reliable solution, several methods have been shown in chapter "Storage solutions".

## *8.7 Kernel software source*

The 3RC main core (kernel) software source is written in bash shell script, the native Linux based system language. The language choice depends on the needs of a robust system infrastructure, minimizing the new software layers addiction. Each software addiction can become an error source.
The whole project, from the PXE architecture to the 3RC main core, is available as open source free software.
The main core source of the 3RC remote controller is provided in Appendix.

## *8.8 Installation*

The 3RC core installation procedure is very simple. After the controller - cluster hosts public key exchange, the service runs as a scheduled script. No root privileges are required to proper operation.
For the Preboot eXecution Environment PXE architecture set up, a brief How To - available online - is recommended.

## *8.9 Configuration*

The 3RC high availability service configuration step entails the set up of the hosts (both virtual and physical) features, such as:
- host type: physical host (PH) or virtual machine (VM);
- hostname;
- physical host load threshold;
- reinstallation flag: only if set to 1 the virtual machine is reinstallable from scratch;
- operating system: distro + architecture (32/64 bit);
- middleware needed by the virtual machine in order to run the proper service.

**hosts.def**

```
  # TYPE HOSTNAME  MAXL INST OS      MW

    PH   physic01   10
    PH   physicNN   10
    VM   virtual1        0    sl4-32  ig_CE
    VM   virtualN        1    sl4-64  ig_WN

  # MAXL: max load
  # INST: VM reinstallable
```

The service time scheduling is provided by

**crontab** or **/etc/cron.d/ha**

```
  # High Availability - every 1 minute
    * * * * *  user  /<PATH>/<HIGH_AVAILABILITY_script>
```

## *8.10 Log*

The 3RC service log reports the date, the virtual machine, the physical host and the performed action:

```
date                  vm - ph1 ph2  last_r               f  action
YYYY-MM-DD/HH:MM:SS -- VM - PH1 PH2 - YYYY-MM-DD/HH:MM:SS [0] ACT/...wait
```

```
     ACTION VM <VM> on PH <PH>2 [from PH <PH1>]
EX: LOG after CRASH VM vrt1
>> Clear history
>> REBOOT   VM vrt1 on PH alfa01
2000-01-01/00:00:00 -- vrt1 - alfa01 alfa01 - 2007-01-01/00:00:00 [1] REBOOT
```

A comprehensive log example will be shown in the next paragraph with a real crash examination.

## *8.11 Operation in a real crash example*

A real crash example is reported.
At 4:e00 AM (night local time) *gridce*, the computing element - the main server - of the SNS-PISA CERN LCG/EGEE Grid environment turns off for an electrical power glitch in the computing room (unknown are the causes of the electrical problem, as often happens).

Figures 8.2, 8.3, 8.4 show respectively the CPU, RAM and Network state of the failed virtual machine *gridce.sns.it.*

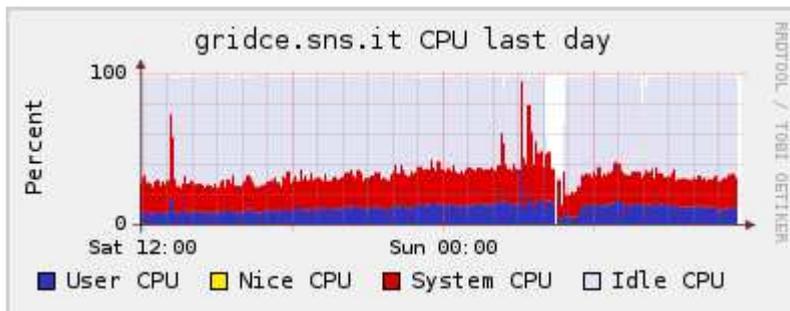

*Fig. 8.2: gridce host CPU monitor*

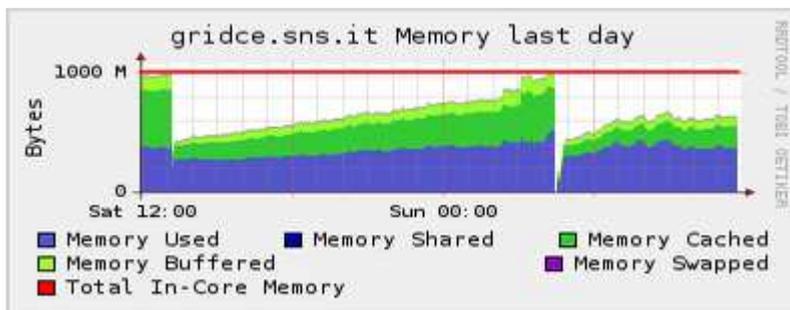

*Fig. 8.3: gridce host Memory monitor*

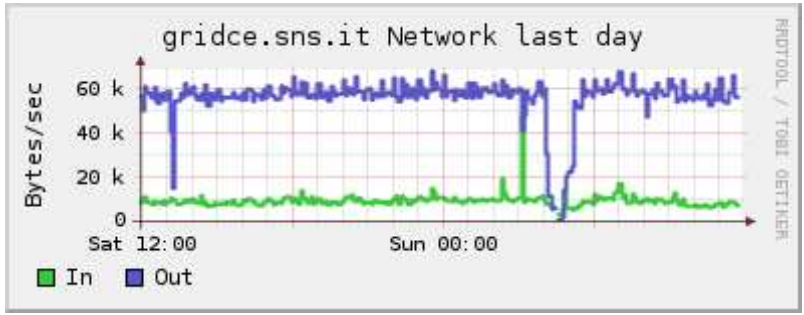

*Fig. 8.4: gridce host Network monitor*

At the electrical power glitch moment the virtual machine *gridce* was hosted by the physical server *alfa01*. At that moment the *alfa01* load was higher than maximum load acceptable (the information is stored in the hosts configuration file). The 3RC choice algorithm searches for another host inside the computing cluster running virtual environment a physical host with a load level lower than threshold: *alfa04*. After an unsuccessful reboot try via ssh (the host is down), the virtual environment involving the crashed virtual machine *gridce* is restarted on *alfa04* physical host.
In less than eight minutes the service is completely restored. The time could have been reduced, by omitting the reboot step, till about 3 - 4 minutes.

Fig. 8.5 shows the *gridce* crashed virtual machine, and *alfa01 alfa04* physical hosts status during the 24 hours including the electrical power glitch.

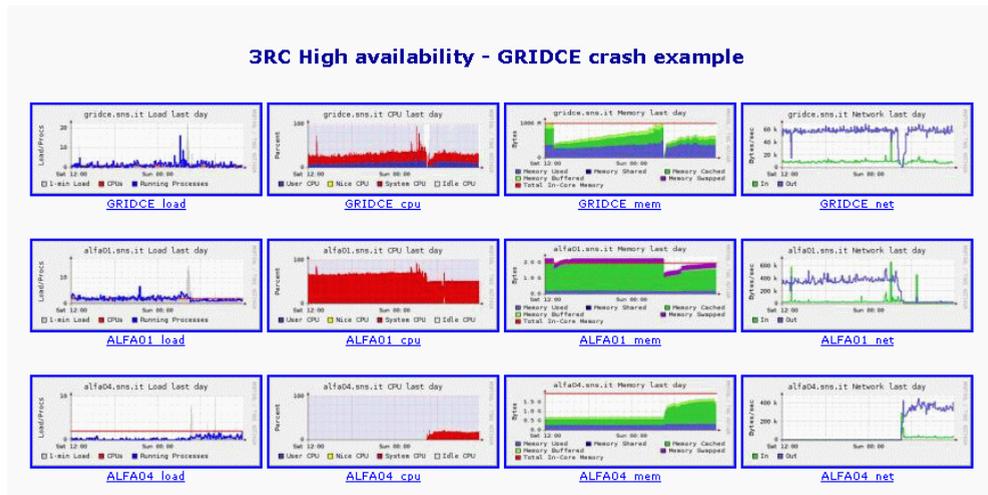

*Fig. 8.5: Crash involved machines status*

The operation log of the *gridce* virtual machine recovery and movement - from physical host *alfa01* to physical host *alfa04* - shows:
- the clear history: no operations have been performed in the last hour, the virtual machine is supposed to work properly;
- virtual machine reboot try via ssh - no successful due to virtual machine unreachability (the host is down);
- 3 minutes time wait;
- physical hosts load level check;
- virtual layer (involving the only *gridce* virtual machine) restart on the available physical host *alfa04;*
- 3 minutes time wait;
- the virtual machine *gridce* becomes responsive - the service is recovered.

```
------------
GRIDCE CRASH
------------

* TRY TO REBOOT

 >> Clear history
 >> REBOOT VM gridce on PH alfa01
 2008-12-14/04:31:01 -- gridce - alfa01 alfa01 - 2008-12-13/13:12:01 [1] REBOOT

* WAIT

 2008-12-14/04:32:01 -- gridce - alfa01 alfa01 - 2008-12-14/04:31:01 [1] ..wait
 2008-12-14/04:33:01 -- gridce - alfa01 alfa01 - 2008-12-14/04:31:01 [1] ..wait
 2008-12-14/04:34:01 -- gridce - alfa01 alfa01 - 2008-12-14/04:31:01 [1] ..wait

* RESTART VM

 >> RESTART VM gridce on PH alfa04 [from OLD PH alfa01]
 2008-12-14/04:35:01 -- gridce - alfa01 alfa04 - 2008-12-14/04:31:01 [2] RESTART

* WAIT

 2008-12-14/04:36:01 -- gridce - alfa01 alfa04 - 2008-12-14/04:35:01 [2] ..wait
 2008-12-14/04:37:01 -- gridce - alfa04 alfa04 - 2008-12-14/04:35:01 [2] ..wait
 2008-12-14/04:38:01 -- gridce - alfa04 alfa04 - 2008-12-14/04:35:01 [2] ..wait

* VM RECOVERED !!!
```

# 9 SPIN-OFF

## 9.1 Host on-demand

As a spin-off of this work, exploiting virtualization and ability to automatically install and configure a host in a few minutes, we could provide a sort of "host on-demand". In a such architecture, the user can ask for a self configured host in terms of processor number, RAM, disk space, operating system and middleware for a given time, with the administrator privileges. At the end of the scheduled time the machines are destroyed.
This way it is possible to optimize the resources sharing across a large number of hosts shared by a large number of users, by simply redistributing the load upon all the available machines [55].

Fig. 9.1 [netdigix.com] shows a typical cloud architecture aimed at providing a host on-demand service.

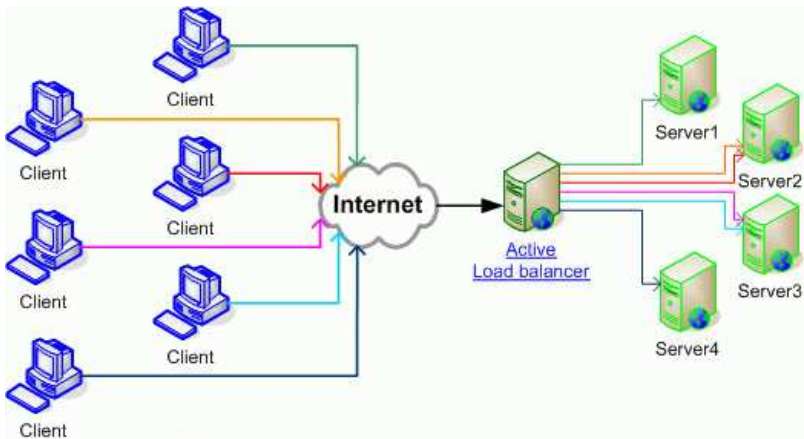

*Fig. 9.1: Host on-demand*

### 9.1.1 Note
This meaning of Host on-demand is not to be confused with the IBM WebSphere Host On-Demand software [56]: a Web browser based access to host applications and data.

## 9.2 Architecture

A front-end web server manages the users requirements: how many servers you need? how many CPUs, RAM, disks? which operating system? how much time for?
The kernel of 3RC project can be used, with some little modifications, to allocate the required disk space on the storage repository and to instantiate the required number of virtual machines.

VMware allows to create a new virtual machine at once, by passing to the system the hardware specifications. Once the virtual machine has been created, using the whole PXE infrastructure is it possible to install the operating system required by the end-user. If the required host is compliant with a known configuration, the system can automatically install the middleware applications too, in order to provide a full operational machine in a few minutes.

This way, in a time from 3 to 5 minutes a user can require and achieve as administrator whatever number of hosts, compatibly with the computing center physical CPU availability and the users policy.

# 10 CONCLUSIONS

Exploiting virtual environment and its features, a high availability service (3RC) has been developed with no additional costs in terms of hardware or software for a computing center, and a very low human effort. This new approach - compliant with the starting requirements - is able to guarantee a relaxed high availability level, with a recovery time of about three minutes for a non destructive crash, and lower than ten minutes if a compromised host needs to be completely reinstalled. The whole Grid data center SNS-PISA is running at the moment in virtual environment, controlled by the 3RC high availability system.

The main difference with respect to the pre-existing solutions is that no operations are required on the computing center hosts in order to guarantee the high availability service. On the other hand the 3RC high availability system does not provide a hot-swap host redundancy. It allows hosts to be restarted or reinstalled in a few minutes, but the operational continuity is not guaranteed.

I would like to refer to a statement by Luigi Picasso, Theoretical Physics Professor at University of Pisa: "It is important to know what a theorem states, but it is probably more important to know what a theorem does not state". As a consequence of this statement, what my project is not intended to be used for? 3RC high availability system does not provide a zero downtime recovery service; it provide a relaxed recovery solution. Therefore, it is not a reliable solution for mission critical applications, such as financial transactions, security certificates management, real time controllers, human health related applications.

As well as the recovery service is able to restart the virtual layer or reinstall the virtual machine from scratch, with a small addition to my software it is possible to implement different recovery solutions. With a scheduled snapshot policy, the system could be restarted, in case of crash, in a previously saved state.
With respect to the storage solutions, an improvement in read/write performance can be achieved using the Logical Volume Manager LVM architecture instead of an image file living on the physical host file system. The use of LVM removes the physical host file system layer, reducing the overhead in the I/O operations.

An interesting future research area will be the study of other virtual environment solutions, such as XEN - not yet mature enough when I started working in virtual environment - or KVM - a very promising module embedded in the Linux kernel.
During the last year, a 3RC similar solution based on open source software Heartbeat over a XEN cluster [57] has been developed. Operationally this solution needs a backup dedicated server or a certain amount of free resources shared over the cluster hosts; this entails a non full exploitation of the computing center resources, and needs a larger number of CPUs than my solution.

Future work can be done in order to improve the system performance or add some extra features. A solution to decrease the time needed for a failed host recovery could be great, maybe not too hard to implement, but would it be really useful?

# APPENDIX

## Kernel source code

```
#!/bin/sh
# HIGH AVAILABILITY SERVICE

# Monitor hosts down, then reboot or restart (reset power/virtual or start on another
Virtual host)

# -------------------------------------------------------------------- #**
#
#  PROC:   High Availability
#  PROG:   Controller, Restart hosts
#
#  AUTH:   FEDERICO CALZOLARI - Scuola Normale Superiore - Pisa [Italy]
#          federico.calzolari@sns.it
#
#  DATE:   2008
#
# -------------------------------------------------------------------- #
```

## Program Flowchart

```
# ---------- PROGRAM FLOWCHART -------------------------------------- #**
#
#  IF DELTA_T < RESTART_TIME
#     NO ACTION (to prevent reinstall after reboot)
#  FI
#  IF DELTA_T > RESTART_TIME
#     IF DELTA_T > 1h
#        flag = 0 (CLEAR history)
#     FI
#     CHECK VMware on Physical host & RESTART VMware if not running
#     IF flag = 0
#        REBOOT    via SSH
#        flag = 1
#     ELSE
#     IF flag = 1
#        RESTART   via VMware
#        flag = 2
#     ELSE
#     IF flag = 2 AND inst = 1
#        REINSTALL via PXE
#        flag = 0
#     FI
#     FI
#     FI
#  FI
#     SET reset time
#
#  - get monitor info from Ganglia                       > HA_mon
#  - extract info (load,heartbeat,status) using sed,awk  > HA_load
#  - get info on host max_load,services                  < ha_host.def
#  - perform CONTROL on
#    - status off  ==> reboot / switch / reinstall
#    - lock file for selected host
#    - if NOT     rebooted ==> reboot
#    - if already rebooted ==> reinstall
#      - call ACTION
#
# -------------------------------------------------------------------- #
```

## Configuration

```
# ---------- CONFIGURATION ----------------------------------------- #**
#
# SETTINGS
#
#   HA_DIR        /var/spool/ha     HA dir: host definition ha_host.def, LOG
#   RESTART_TIME  600               max time needed for host restart
#   DEBUG         0,1               debug  mode (0:OFF 1:ON)
#   ACTION        0,1               action mode (0:OFF 1:ON) 0:FAKE, NO action taken
#
#
# FILES
#
#   /usr/bin/ha                     main High Availability controller/action
#   HA_DIR/host.def                 host definition (TYPE HOSTNAME OS MW)
#   HA_DIR/log                      log
#   HA_DIR/vm/<VM>                  virtual machines hosted by, restarted at, flag
#
#   ---------
# > hosts.def
#   ---------
#   # TYPE HOSTNAME   MAXL INST OS      MW
#     PH   physic01    10
#     PH   physicNN    10
#     VM   virtual1         0   sl4-32  ig_CE
#     VM   virtualN         1   sl4-64  ig_WN
#   # VM   extra           0/1  [# NOT included in HA service]
#   # MAXL: max load
#   # INST: VM reinstallable
#
#   -----
# > vm/VM
#   -----
#   PH_owner_VM last_restart FLAG
#
#
# SCHEDULER: crontab
#   > /etc/cron.d/ha
#     # High Availability - every 1 minute
#     * * * * *   root  /bin/ha
#   OR
#   > /etc/cron.d/ha                cron scheduler
#
# ------------------------------------------------------------------ #
```

## Install HowTo

```
# ---------- INSTALL ----------------------------------------------- #**
#
# INSTALL [as root/user]
#
#   mkdir -p       /var/spool/ha/vm
#   chmod -R ugo+rw /var/spool/ha/
#
#   SET:
#     /usr/bin/ha
#     /var/spool/ha/
#     /var/spool/ha/vm
#     /var/spool/ha/vm/<VM>
#
# ------------------------------------------------------------------ #
```

## Events Log

```
# ---------- LOG ---------------------------------------------------- #**
#
# LOG
#
# date                 vm - ph1 ph2   last_r              f  action
# YYYY-MM-DD/HH:MM:SS -- VM - PH1 PH2 - YYYY-MM-DD/HH:MM:SS [0] ACT/...wait
#      ACTION VM <VM> on PH <PH>2 [from PH <PH1>
#
# EX: LOG after CRASH VM vrt1
#
# >> Clear history
# >> REBOOT  VM vrt1 on PH alfa01
# 2000-01-01/00:00:00 -- vrt1 - alfa01 alfa01 - 2007-01-01/00:00:00 [1] REBOOT
# 2000-01-01/00:01:00 -- vrt1 - alfa01 alfa01 - 2008-01-01/00:00:00 [1] ..wait
# 2000-01-01/00:02:00 -- vrt1 - alfa01 alfa01 - 2008-01-01/00:00:00 [1] ..wait
# >> RESTART VM vrt1 on PH alfa02 [from OLD PH alfa01]
# 2000-01-01/00:03:00 -- vrt1 - alfa01 alfa02 - 2008-01-01/00:00:00 [2] RESTART
# 2000-01-01/00:04:00 -- vrt1 - alfa02 alfa02 - 2008-01-01/00:00:00 [2] ..wait
# 2000-01-01/00:05:00 -- vrt1 - alfa02 alfa02 - 2008-01-01/00:00:00 [2] ..wait
# >> REINSTALL VM vrt1 on PH alfa03 [from OLD PH alfa02]
# 2000-01-01/00:06:00 -- vrt1 - alfa02 alfa03 - 2000-01-01/00:03:00 [0] REINSTALL
# 2000-01-01/00:07:00 -- vrt1 - alfa03 alfa03 - 2000-01-01/00:03:00 [0] ...wait
# 2000-01-01/00:18:00 -- vrt1 - alfa03 alfa03 - 2000-01-01/00:03:00 [0] ...wait
# 2000-01-01/00:09:00 -- vrt1 - alfa03 alfa03 - 2000-01-01/00:03:00 [0] ...wait
# 2000-01-01/00:10:00 -- vrt1 - alfa03 alfa03 - 2000-01-01/00:03:00 [0] ...wait
# [VM RECOVERED]
#
# ---------------------------------------------------------------------- #
```

## Test procedure

```
# ---------- TEST ---------------------------------------------------- #**
#
# set ACTION=0
#
# echo "PH DATE/TIME FLAG" > /var/spool/ha/vm/<VM>
# EX:  echo "alfa01 2008-01-01/00:00:00 0" > /var/spool/ha/vm/vrt1
#
# . ha
#
# ---------------------------------------------------------------------- #
```

## Settings

```
# ---------- SETTINGS ---------------------------------------- #**
#
# -------- #
# SETTINGS #
# -------- #
#
  set -o nounset
   DEBUG=0              ## DEBUG  mode [0/1]
   ACTION=1             ## ACTION mode [0/1]
   RESTART_TIME=600        ## min uptime (sec) to be sure that the host has really restarted

   # HA path
   HA_DIR=/var/spool/ha
```

## Functions

```
# ---------- FUNCTIONs START --------------------------------- #**
#
#    function DEBUG
     function DBG() { [ $DEBUG = 1 ] && echo -e "   @@ DEBUG: $1" >&2; }

#    function LOG($vm $ph $ph2 $LAST_R $FLAG $ACT)
     function LOG() {

            vm=$1
           ph1=$2
           ph2=$3
        LAST_R=$4
          FLAG=$5
           ACT=$6

        #cho -e "# --------------------------- #" | tee -a $HA_DIR/log
        echo -e "`date '+%F/%T'` -- $vm - $ph1 $ph2 - $LAST_R [$FLAG] $ACT" | tee -a $HA_DIR/log
     }

#    function CHECK_LOAD
     function CHECK_LOAD() {
         ph=$1
         maxl=`grep $ph $HA_DIR/hosts.def | awk '{print $3}'`
         load=`grep $ph /tmp/HA_load      | awk -F";" '{print $2}'`
         maxl=`echo $maxl *100 | bc -l    | awk -F"." '{print $1}'`
         load=`echo $load *100 | bc -l    | awk -F"." '{print $1}'`
         beat=`grep $ph /tmp/HA_load      | awk -F";" '{print $3}'`
     }
#
# ---------- FUNCTIONs END ----------------------------------- #
```

## Actions

```
# ---------- ACTION FUNCTIONs-------------------------------- #**
#
    function VMrestart() {
         # RESTART VM via VMware [FUNCTION]
            vm=$1
           ph1=$2
           ph2=$3
           ssh  root@$ph1  "nn=\`vmware-vim-cmd vmsvc/getallvms | grep $vm | awk '{print \$1}'\`;
                            # echo VMid: \$nn;
                            vmware-vim-cmd vmsvc/power.off \$nn > /dev/null 2>&1;
```

```
                                [ `ps auxww | grep -v grep | grep -i $vm | wc -l` != 0 ] &&
ps auxww | grep -v grep | grep -i $vm | awk '{print \$2}' | xargs kill -9;
                                rm -f /opt/vmware/$vm/$vm.vmdk.lck/*;
                                " &
            sleep 10
            ssh  root@$ph2  "nn=\`vmware-vim-cmd  vmsvc/getallvms  |  grep  $vm  |  awk
'{print \$1}'\`;
                                # echo VMid: \$nn;
                                vmware-vim-cmd vmsvc/power.off \$nn > /dev/null 2>&1; sleep
10;
                                [ `ps auxww | grep -v grep | grep -i $vm | wc -l` != 0 ] &&
ps auxww | grep -v grep | grep -i $vm | awk '{print \$2}' | xargs kill -9;
                                rm -f /opt/vmware/$vm/$vm.vmdk.lck/*;
                                vmware-vim-cmd vmsvc/power.on  \$nn > /dev/null 2>&1;
                                " &
    }

    function REBOOT() {
            # REBOOT VM via SSH
             vm=$1
            ph1=$2
            ph2=$3
            echo ">> REBOOT  VM $vm on PH $ph1" | tee -a $HA_DIR/log
            [ $ACTION = 1 ] && ssh root@$vm "reboot" &
    }

    function RESTART() {
            # RESTART VM via VMware
             vm=$1
            ph1=$2
            ph2=$3
            echo ">> RESTART VM $vm on PH $ph2 [from OLD PH $ph1]"   | tee -a
$HA_DIR/log
            [ $ACTION = 1 ] && VMrestart $vm $ph1 $ph2
    }

    function REINSTALL() {
            # REINSTALL VM via PXE
             vm=$1
            ph1=$2
            ph2=$3
            echo ">> REINSTALL VM $vm on PH $ph2 [from OLD PH $ph1]" | tee -a
$HA_DIR/log
            # SET PXE environment
            IP2HEX=`perl /var/lib/tftpboot/pxelinux.cfg/iphex.pl $vm | awk '{print
$3}'`
            OS=`grep $vm $HA_DIR/hosts.def | awk '{print $4}'`
            [ $ACTION = 1 ] && ln -fs $OS /var/lib/tftpboot/pxelinux.cfg/$IP2HEX
            [ $ACTION = 1 ] && VMrestart $vm $ph1 $ph2
    }
#
# ---------- END ACTION FUNCTIONS ----------------------------------- #
```

## Main program

```
# ---------- PROGRAM ---------------------------------------------- #**
```

## Ganglia monitor

```
# --- MONITOR ------------------------------------------------------ #
#
# GET MONITOR INFO
  wget -q http://bibe.sns.it/ganglia/?hc=1\&p=3\&c=cluster -O /tmp/HA_mon

# EXTRACT INFO
  grep   -A2   "<td><a   href"   /tmp/HA_mon   |   sed   -e   's/.*">//g;   s/<\/a>.*//g;
s/.*<small>//g;  s/<\/small>.*//g;    s/heartbeat  s/heartbeat  9999s/g;  s/.*beat  //g;
s/s<\/font.*//g;'  |  sed  'N;s/\n/;/g'  |  sed  'N;s/\n/;/g'  |  sed  -e  's/--//g'  >
/tmp/HA_load
#
# ------------------------------------------------------------------ #
```

## Check control

```
# --- CONTROL ------------------------------------------------------ #
#
# CHECK OFFLINE MACHINEs

  ## vm: virtual machine
  ## ph: physical host

  for hh in `grep 9999 /tmp/HA_load | awk -F";" '{print $1}' | awk -F"." '{print $1}'`;
do
   # check if host is a VM
     if [ `grep $hh $HA_DIR/hosts.def | grep -v "#" | grep "VM" | wc -l` != 0 ]; then
        vm=$hh
        ph=`awk '{print $1}' $HA_DIR/vm/$vm`

       # ORIG PH
         ph1=$ph

       # set PH to restart VM --> ph2
       # check if vm physical host owner is up and its load is lower than MAXL
         CHECK_LOAD $ph
         if [ $beat != 9999 ] && [ $load -le $maxl ]; then
          # the SAME PH $ph2 is available
            ph2=$ph

         else
          # #ANOTHER PH $ph2 selected
          # search for lower load PH
            for pbk in `grep "PH" $HA_DIR/hosts.def | grep -v "#" | grep -v "$ph" | awk
'{print $2}'`; do
              CHECK_LOAD $pbk
              if [ $beat != 9999 ] && [ $load -le $maxl ]; then
                 ph2=$pbk
              fi
            done
         fi
         #echo "PH: $ph2 selected to restart" | tee -a $HA_DIR/log

      # check last restart --> DELTA_T
        NOW=`date '+%F/%T'`
        NOW2=`date +%s`
        LAST_R=`awk '{print $2}' $HA_DIR/vm/$vm`
```

```
            LAST_R2=`date -d "\`awk '{print $2}' $HA_DIR/vm/$vm | awk -F"/" '{print $1 " " $2}'`" '+%s'`
            FLAG=`awk '{print $3}' $HA_DIR/vm/$vm`
            let DELTA_T=$NOW2-$LAST_R2

        # SET RESTART_TIME depending on last action
            if [ $FLAG = 1 ]; then RESTART_TIME=200; fi  # after reboot
            if [ $FLAG = 2 ]; then RESTART_TIME=200; fi  # after restart
            if [ $FLAG = 0 ]; then RESTART_TIME=600; fi  # after reinstall OR clear
#
# ------------------------------------------------------------------ #
```

## Operation

```
# --- ACTION --------------------------------------------------- #
#
        # IF LAST RESTART <  RESTART_TIME
          if [ $DELTA_T -le $RESTART_TIME ]; then
             ACT="...wait"
             LOG $vm $ph1 $ph1 $LAST_R $FLAG $ACT
          fi

        # IF LAST RESTART >  RESTART_TIME
          if [ $DELTA_T -gt $RESTART_TIME ]; then
             ACT="ACT"

          # LAST RESTART > 1H  ==> Clear history
            if [ $DELTA_T -ge 3600 ]; then
                echo ">> Clear history" | tee -a $HA_DIR/log
                FLAG="0"
                echo "$ph2 $LAST_R $FLAG" > $HA_DIR/vm/$vm
            fi

          # CHECK/Restart VMware on Physical host
            [ $ACTION = 1 ] && \
            ssh root@$ph2 "if [ `ps auxw | grep vmware | grep -v 'grep' | wc -l` = 0 ]; then
                             /etc/init.d/vmware restart;
                          fi"

          # ------ #
          # ACTION #
          # ------ #
            if [ $FLAG = 0 ]; then
                REBOOT    $vm $ph1 $ph1
                ACT="REBOOT"
                FLAG="1"
            else
            if [ $FLAG = 1 ]; then
                RESTART   $vm $ph1 $ph2
                ACT="RESTART"
                FLAG="2"
            else
            if [ $FLAG = 2 ]; then
             # chech if VM is one which can be reinstalled
                INST=`grep $vm $HA_DIR/hosts.def | awk '{print $3}'`
                [ $INST = 1 ] && REINSTALL $vm $ph1 $ph2
                [ $INST = 1 ] && ACT="REINSTALL"
                FLAG="0"
            fi
            fi
            fi

          # LOG write
            LOG $vm $ph1 $ph2 $LAST_R $FLAG $ACT

          # SET reset time
            echo   "$ph2 $NOW $FLAG" > $HA_DIR/vm/$vm
```

```
            fi
#
# ---------------------------------------------------------------- #
    fi
  done
#
# ---------- END PROGRAM ---------------------------------------- #
```